\documentclass{aa}  

\usepackage{natbib}
\bibpunct{(}{)}{;}{a}{}{,} 
\usepackage{enumitem}
\usepackage{amsmath}
\usepackage{blindtext}
\usepackage{graphicx}
\usepackage{txfonts}
\usepackage{xcolor}

\newcommand{\cmfast}{{\texttt{21cmFAST}}}
\newcommand{\disperse}{{\texttt{DisPerSE}}}
\newcommand{\emma}{{\texttt{EMMA}}}
\newcommand{\amber}{{\texttt{AMBER}}}

\DeclareMathOperator\erf{erf}
\DeclareMathOperator\erfc{erfc}
\newcommand{\textvec}[1]{
  \left(\begin{matrix}#1\end{matrix}\right)%
}

\begin{document}

   \title{Topology of Reionisation times: concepts, measurements and comparisons to gaussian random field predictions}

   \author{Emilie Thélie
          \and
          Dominique Aubert
          \and
          Nicolas Gillet
          \and
          Julien Hiegel
          \and
          Pierre Ocvirk
          }

   \institute{Université de Strasbourg, CNRS, Observatoire astronomique de Strasbourg, UMR 7550, F-67000 Strasbourg, France\\
              \email{emilie.thelie@astro.unistra.fr}}

   \date{Received ..., accepted ...}

 
  \abstract
   {In the next decade, radio telescopes like the Square Kilometer Array (SKA) will explore the Universe at high redshift, and particularly during the Epoch of Reionisation (EoR). The first structures emerged during this epoch, and their radiations have reionised the previously cold and neutral gas of the Universe creating ionised bubbles that percolate at the end of the EoR ($z\sim6)$. SKA will produce 2D images of the distribution of the neutral gas at many redshifts, pushing us to develop tools and simulations to understand its properties.}
   {This paper aims at measuring topological statistics of the EoR in the so-called ‘reionisation time' fields from both cosmological and semi-analytical simulations. This field informs us about the time of reionisation of the gas at each position, is used to probe the inhomogeneities of reionisation histories and can possibly be extracted from 21 cm maps. We also compare these measurements with analytical predictions obtained within the gaussian random field (GRF) theory.}
   {The gaussian random fields theory allows us to compute many statistics of a field: probability distribution functions (PDFs) of the field or its gradient, isocontour length, critical point distributions, and skeleton length. We compare these theoretical predictions to measurements made on reionisation times fields extracted from an $\emma$ and a $\cmfast$ simulations at 1 a cMpc/h resolution. We also compared our results to GRFs generated from the fitted power spectra of the simulation maps.}
   {Both $\emma$ and $\cmfast$ reionisation times fields ($t_{\text{reion}}(\Vec r)$) are close to be gaussian fields, in contrast with the 21 cm, density or ionisation fraction that are all proven to be non-gaussian. Only accelerating ionisation fronts at the end of the EoR seem to be a cause of small non-gaussianities in $t_{\text{reion}}(\Vec r)$. Overall, this topological description of reionisation times provides a new quantitative and reproducible way to characterize EoR scenario. Under the assumption of gaussian random fields, it enables the generation of reionisation models with their propagation, percolation or seeds statistics from the sole knowledge of the reionisation times power spectrum. Conversely, these topological statistics provide means to constrain the power spectrum properties and by extension the physics that drive the propagation of radiation.}
  {}


   \maketitle
%
\section{Introduction}
\label{sec:introduction}

The Epoch of Reionisation (EoR) sees the birth of stars and galaxies. The first sources of radiation appear during the EoR while emitting photons that reionise the cosmic gas and create HII ‘bubbles' around galaxies. These bubbles eventually percolate near the end of the EoR at $z=5.3-6$ \citep{Barkana2001, Dayal2018, Kulkarni2019, Wise2019}. This epoch marks the transition from a totally cold and neutral Universe gas to today's warmer and ionised one.

The evolving geometry of the EoR is widely investigated in the literature in order to understand physical processes, such as the growth of structures, the geometry of the ionised or neutral bubbles, or the percolation. 
Many works focus on the geometry of the ionised/neutral bubbles and on percolation with Minkowski functionals (or derived statistics, such as the Euler characteristic, the genus, or the shapefinders; see \citet{Gleser2006,Lee2008,Friedrich2011,Hong2014,Yoshiura2017,Chen2019,Pathak2022}), with the triangle correlation function \citep{Gorce2019}, or with the Morse theory and persistent homology \citep{Thelie2022}.
Other studies extract the size and shape of the ionised bubbles thanks to the contour Minkowski tensor  \citep{Kapahtia2018,Kapahtia2019,Kapahtia2021}. 
Counting numbers of 3D structures of a field (isolated objects like peaks, tunnels or voids) can be done using the Betti numbers \citep{Kapahtia2019,Kapahtia2021,Giri2020b,Bianco2021,Elbers2022}.
The size of the ionised or neutral bubbles are also investigated with methods such as the friend-of-friend algorithm \citep{Iliev2006,Friedrich2011,Lin2016,Giri2018,Giri2019a}, the spherical average method \citep{Zahn2007,Friedrich2011,Lin2016,Giri2018}, the mean free path method \citep{Mesinger2007,Lin2016,Giri2018,Giri2019a,Bianco2021}, or the granulometry method \citep{Kakiichi2017,Busch2020}.
Besides, the low-frequency component of the Square Kilometre Array radio interferometer\footnote{https://www.skatelescope.org} (SKA-Low; see e.g. \citet{Mellema2013}) will produce 2D tomographic images of the 21 cm HI emission at many redshifts during the EoR. There are therefore many studies on this signal spatial structure with the 21 cm power spectrum for example \citep{Zaldarriaga2004,Furlanetto2004,McQuinn2006,Bowman2006,Lidz2008,Iliev2012,Mesinger2013,Pober2014,Greig2015,Liu2016,Greig2017,Kern2017,Park2019,Pagano2020,Gazagnes2021}, or the 21 cm bispectrum \citep{Hutter2020}.

\paragraph{Reionisation times field: definition and motivation}

\begin{figure}
   \centering
   \includegraphics[width=0.5\textwidth]{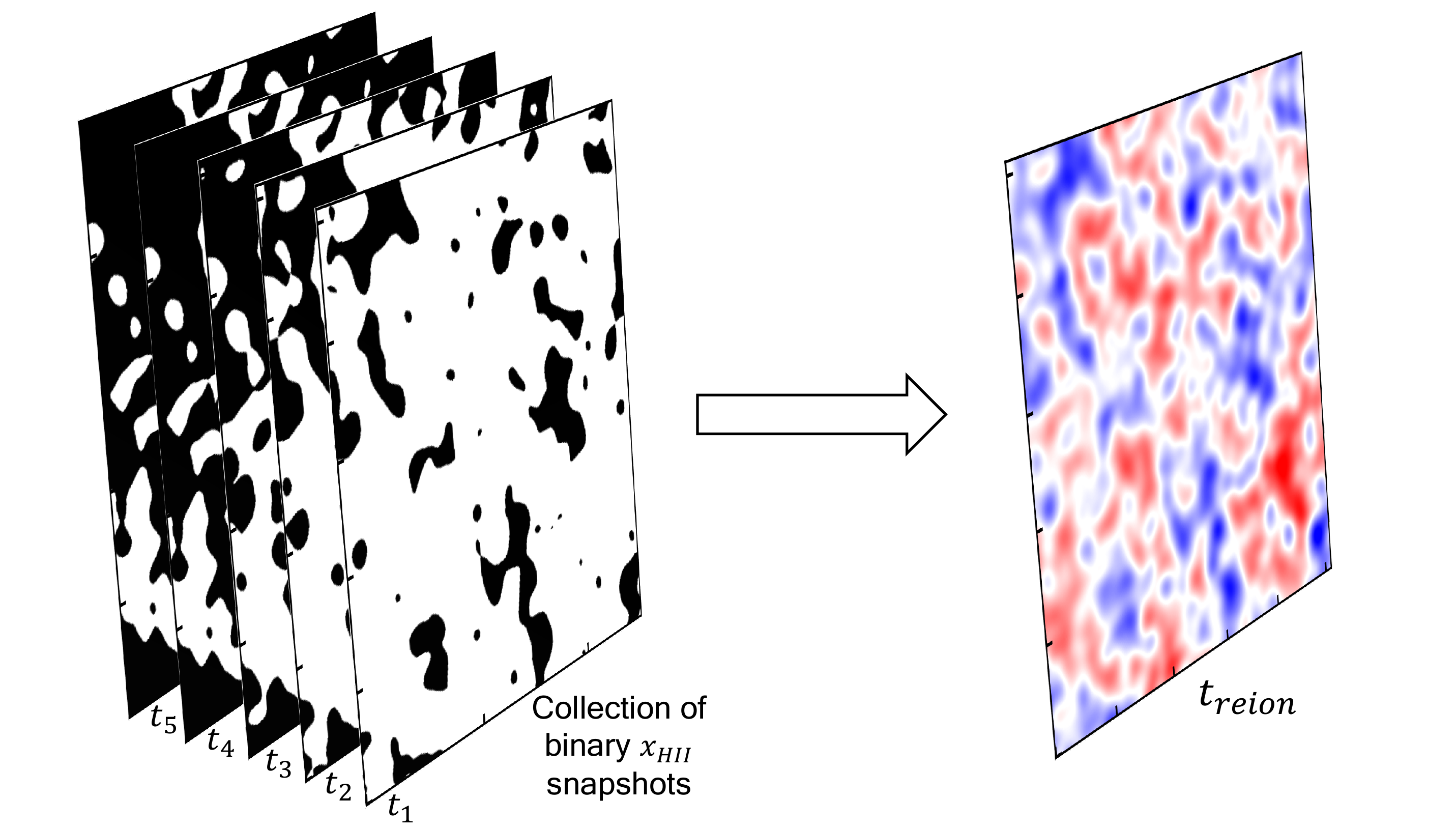}
    \caption{Schematic summarising the use of the reionisation times field (on the right): it allows us to use only one field instead of a series of many snapshots of binary ionised fraction (on the left) for example.}
    \label{fig:treion_scheme}
\end{figure}

\begin{figure*}[ht]
   \centering
   \includegraphics[width=1\textwidth]{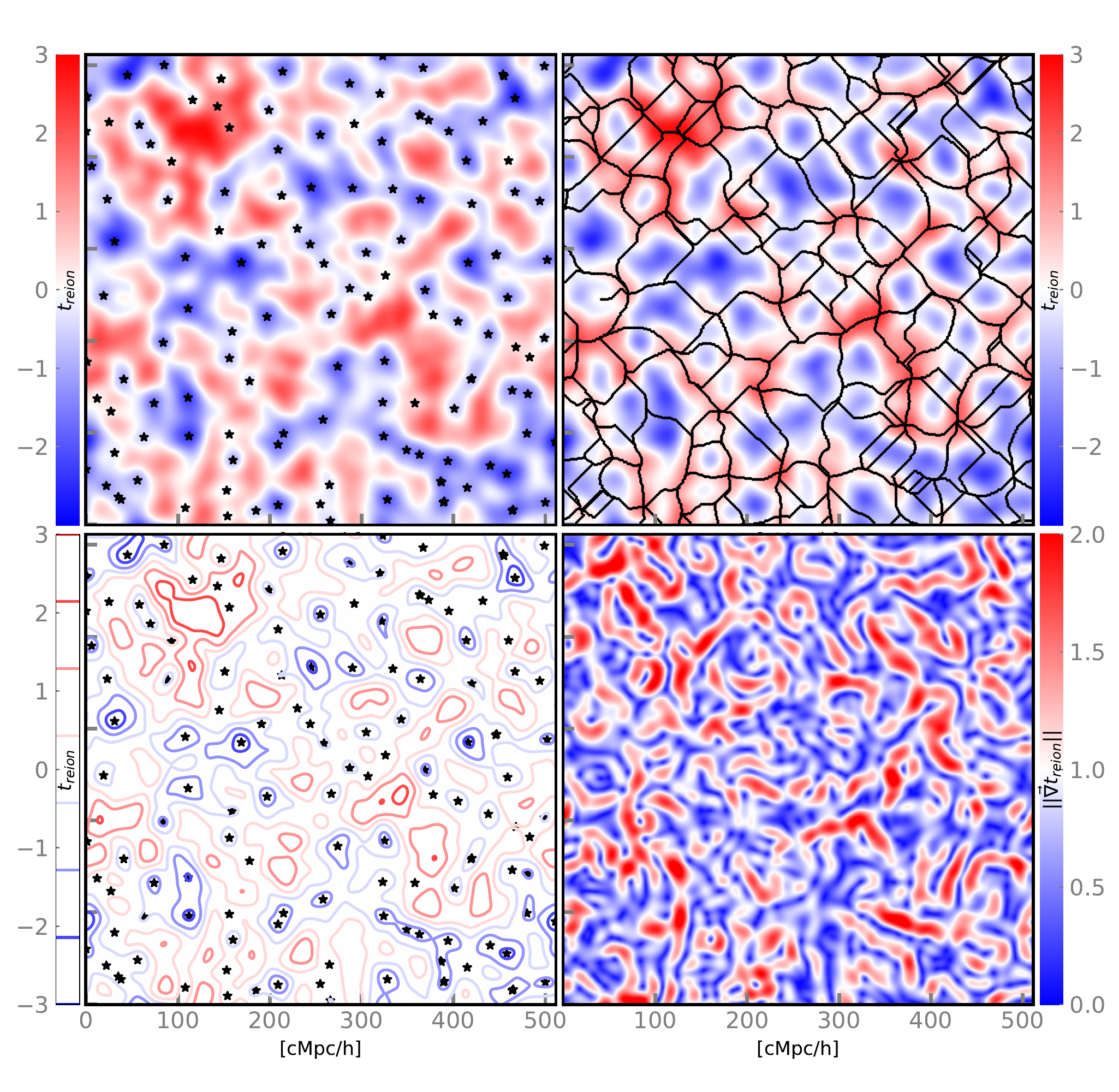}
    \caption{2D slices related to the $\emma$ reionisation times field that is smoothed with a gaussian kernel of standard deviation of $R_f=6$. On the top left panel, $t_{\text{reion}}(\Vec r)$ is shown with its minima (black stars). On the top right panel, $t_{\text{reion}}(\Vec r)$ is shown with its skeleton (black lines). On the bottom left panel, the isocontours of $t_{\text{reion}}(\Vec r)$ are represented, with its minima again in black stars. Here, the reionisation times field is normalised to put its mean at 0, and its standard deviation at 1. The bottom right panel shows the norm of the gradient of $t_{\text{reion}}(\Vec r)$.}
    \label{fig:Map_EMMA_treion_project}
\end{figure*}
Our works focus on the reionisation times (or redshifts) map. This field is generated by cosmological simulations/models (e.g. $\emma$ and the $\cmfast$ semi-analytical code): it corresponds to the time at which each position of the simulation box is considered to be reionised, that is when the ionisation fraction exceeds a 50\% threshold, as follows: 
\begin{equation}
    t_{\text{reion}}^*(\vec r)=t(\vec r,x_{HII}=0.5),
\end{equation}
where $\vec r$ is the position, and $x_{HII}$ the ionised fraction. $t^*_{\text{reion}}$ is measured from the Big Bang, meaning that the cosmic gas is almost entirely reionised approximately at $t^*_{\text{reion}} \sim 1$ Gyr. As shown in Fig. \ref{fig:treion_scheme} and thanks to $t_{\text{reion}}(\Vec r)$, we compress the information about the evolution of $x_{HII}$ in a single field instead of using a collection of snapshots. In the reionisation times map, blue regions correspond to those where the gas reionises the first, whereas the red ones are the last ones to reionise. 
We focus on 2D $t_{\text{reion}}(\Vec r)$ maps to study the EoR on the sky as it will be observed (with the upcoming SKA 2D images for instance).

The field $t_{\text{reion}}(\Vec r)$ holds both spatial and temporal information on the reionisation scenario and is thus often used to characterize or compare the evolving structure of the reionisation provided by models. For example, it can be used to measure how fast and along which directions ionising radiation propagate from sources \citep{Deparis2019, Thelie2022}. Recently, it has also been used to generate efficiently models of the reionisation \citep{Trac2021}. $t_{\text{reion}}(\Vec r)$ is also valuable to investigate local variations of the reionisation scenario \citep{Trac2008,Battaglia2013,Aubert2018,Zhu2019,Sorce2022} and the consequences of an inhomogeneous reionisation. These local modulations of reionisation histories could possibly manifest themselves in the star formation histories of low mass galaxies \citep{Ocvirk2020} or their spatial distribution \citep{Ocvirk2011}. 
$t_{\text{reion}}$ is thus a versatile descriptor of models and in this paper we propose to revisit its study in a more general manner. In particular we show how \textit{the topological study} of this field can unravel many properties of the summarized reionisation scenario, in a physically meaningful, quantitative and reproducible way.

However, we also claim that this field, and the study of its topology, is not only useful in the strict and limited scope of reionisation models but also in the context of future observations. Indeed, in the next decade, radio-telescopes like SKA will map the intergalactic medium (IGM) during the EoR thanks to the 21 cm radiation coming from neutral hydrogen atoms (e.g. \cite{Koopmans2015}). 
21 cm lightcones along the line of sight will contain a wealth of information on the evolving reionisation state of the IGM and on the underlying matter density, ionisation fraction of the gas or thermal state. In Hiegel et al. (in preparation), we found that it is also possible to reconstruct 2D reionisation times maps from 2D 21 cm images thanks to a convolutional neural network.
Hence $t_{\text{reion}}(\Vec r)$ could possibly be extracted from observations (even though small structures are smoothed out) thus granting access to the evolution of the reionisation in the transverse plane, in complement to line-of-sight studies. More details about these reconstructions are given in App. \ref{app:hiegel2022}. 
Down the road, we intend to use the framework described in the present paper on the reionisation times field reconstructed from 21 cm observations.

\paragraph{Topology of the reionisation times field}

 The topological features of  $t_{\text{reion}}(\Vec r)$ can be described by  analogy with a mountainous landscape \citep{Gay2011}. Consider a mountainous landscape as a 2D field: for each 2D position of the space, the altitude is the value of the field. The mountain peaks are the maxima of the field, the bottoms of the valley are the minima, and the mountain passes are the saddle points. The skeleton of this field corresponds to all the ridge lines, which join each pass (i.e. saddle point) to peaks (i.e. maxima) and are the lines having the least slope. The skeleton forms a connected network in all the space. 
With $t_{\text{reion}}(\Vec r)$, many geometrical quantities can be interpreted physically to describe the evolution of the ionised and neutral gas during the EoR.
Recently, in \citet{Thelie2022}, we have studied 3D topological properties of the EoR, such as the shape, size and orientation of ‘peak patches of reionisation'. Here, we go further and work on a large set of geometrical properties of the 2D reionisation times fields $t_{\text{reion}}(\Vec r)$, as shown in Fig. \ref{fig:Map_EMMA_treion_project}.

\begin{description}
    \item[\textit{Filling factor and PDF of the field values.}] A widely used statistic when studying the EoR is the fraction of ionised volume $Q_{\text{HII}}$. We can measure it with the reionisation times field as it is directly the cumulated distribution function of the $t_{\text{reion}}(\Vec r)$ values (i.e. the number of cells that has a lower reionisation time than a given threshold). In other words, it is the reionisation history: it contains information about the timing of reionisation, as well as its global evolution.
    
    \item[\textit{PDF of the gradient norm field.}] Moreover, we can compute the first derivative of the reionisation times field, and extract the norm of its gradients. It corresponds to the time interval on which the gas is reionised in each cell of the simulation boxes ($\sim \Delta t/\Delta x$), which is equivalent to the inverse of a velocity field. An example of this gradients norm field is shown on the bottom right panel of Fig. \ref{fig:Map_EMMA_treion_project}. The analysis of this field allows us to study the ‘velocity' of the radiation fronts \citep{Deparis2019}.
    
    \item[\textit{Isocontours length.}] The isocontours of $t_{\text{reion}}(\Vec r)$ locate the regions reached at a given time by the HII bubbles. Their length is interesting because it contains information about the growth of the ionised bubbles and the decrease in size of the last neutral bubbles. With the bottom left panel of Fig. \ref{fig:Map_EMMA_treion_project}, we see the first ionised bubbles with the darkest blue contours, their growth with the lighter blue contours, and their fusion when the blue contours merge. We also see the last neutral regions being ionised with the red contours. Here, we have an insight on the percolation process of the EoR. 
    
    \item[\textit{{Reionisation seeds count.}}] We can extract the critical points of the reionisation times field, and in particular its minima. They correspond to the sources of radiation (first zones that reionise). They are shown on the example field of Fig. \ref{fig:Map_EMMA_treion_project} with the black stars on the left panels. They are as expected within the bluest zones, which reionise the first, as well as in the middle of the first blue isocontours. Their distribution as a function of the time allows us to know the time of appearance of these reionisation seeds. For instance, we can infer the moment when the maximum number of sources lights up. This distribution should also correlate to the star formation rate.
    
    \item[\textit{Reionisation patches.}] We can extract the voids patches from the reionisation times field (or the peak patches from the reionisation redshifts field). They contain all of the cells that are linked to the $t_{\text{reion}}(\Vec r)$ minima by a negative gradient. Thanks to them, we can study the extent of the radiative influence of a reionisation seed with size distributions. Their shape and orientation with respect to the density filaments informs us about the direction of propagation of the reionisation fronts. As we studied these patches in \citet{Thelie2022}, we will not focus on them in this paper. 
    
    \item[\textit{Distribution of the skeleton lengths.}] We also calculate the distribution of the length of the skeleton of $t_{\text{reion}}(\Vec r)$. An example skeleton is shown with the black lines on the top right panel of Fig. \ref{fig:Map_EMMA_treion_project}. As explained before, they connect the maxima of a field by passing through its saddle points as ridge lines. The skeleton of the reionisation times field $t_{\text{reion}}(\Vec r)$ physically corresponds to the front lines between the propagating radiations that come from the reionisation seeds. It indicates therefore to what extent the photons can propagate from a source and ionise the medium before reaching an opposite ionising front coming from another source: it is the percolation lines between patches of reionisation. The skeleton length distribution with respect to the time tells us about the length of merging radiation fronts at a given time.
\end{description}

In this work, we study all of these properties (except the reionisation patches) through measurements in $t_{\text{reion}}(\Vec r)$ maps that are extracted from simulations obtained with the $\emma$ cosmological code and the $\cmfast$ semi-analytical code.

\paragraph{Gaussianity of the reionisation times field}

We also analyse the gaussianity of the reionisation times field ($t_{\text{reion}}(\Vec r)$), and therefore of the reionisation process.
The EoR is known to be ruled by strongly non-gaussian phenomena that are probed in different ways in the literature. Many studies look at this non-gaussian nature of the EoR directly through the 21 cm probability distribution function (PDF), using sometimes skewness, kurtosis, and quantiles analyses \citep{Mellema2006,Ichikawa2010,Dixon2016,Ross2019,Banet2021}. Other studies use higher order statistics, like the bispectrum or trispectrum \citep{Majumdar2018,Shaw2020}.
The density or ionisation fraction fields are also studied for their non-gaussian features \citep{Iliev2006}. The non-gaussianities of the EoR are generally said to be due to the non-linear structure formation \citep{Bernardeau2002}, present at all scales, and are more important the further the reionisation process goes. Moreover, some studies show that non-gaussianities are increased within the HII bubbles due to high ionisation and high densities \citep{Iliev2006,Dixon2016,Majumdar2018}, which could be due to an inside-out process of ionisation \citep{Iliev2006}. \citet{Ross2019} also explains that the quasi-stellar objects (QSOs) and the X-ray heating can be another cause of non-gaussianity. Overall, understanding the source of non-gaussianities will help us to the global comprehension of the EoR physical processes, as well as to put constraints on the reionisation parameters \citep{Shaw2020,Greig2022}.

We compare in this study the measured statistics on $t_{\text{reion}}(\Vec r)$ mentioned above to GRF theory predictions. This theory allows us to compute statistics of gaussian distributed field \citep{Rice1944,Longuet1957,Doroshkevich1970,BBKS,Hamilton1986,Pogosyan2009b,Pogosyan2009,Pichon2010,Gay2012,Cadiou2020} or weakly non-gaussian fields \citep{Matsubara2003,Pogosyan2011,Gay2012,Cadiou2020,Matsubara2021}.
\citet{Rice1944} firstly introduces the GRF theory to extract statistics from the one dimensional random noise of electronic devices. \citet{Longuet1957} uses later the same theory, but this time on random waves on 2D surfaces. It is only later, that \citet{BBKS} use the GRF theory in astrophysics in order to study the 3D structure formation in a cosmological context with the sole hypothesis of a power law power spectrum. More recently, \citet{Gay2012} extract many statistics of 2D or 3D cosmological fields thanks to this theory, counting for example the peaks on made-up cosmic microwave background (CMB) maps to study their non-gaussianities. 
Some widely used statistics in the context of the EoR can be analytically calculated thanks to the GRF theory, such as Minkowski functionals, the derived genus or Euler characteristics, and Betti numbers \citep{Schmalzing1998,Matsubara2003,Lee2008,Gay2012,Kapahtia2019,Matsubara2021}. For example, \citet{Lee2008} compute the genus of the neutral hydrogen field $x_{HII}$ to study the evolution of the EoR through many phases. 

Here, we will compare the statistics of the EoR that we measure on $t_{\text{reion}}(\Vec r)$ with the analytic predictions of the GRF theory, for the first time. 
The advantage of GRFs is that all the field information is compressed within their power spectrum: we thus envision the prospect of summarising the timing and the evolution of the EoR with $t_{\text{reion}}(\Vec r)$ or its power spectrum. 
A gaussian  $t_{\text{reion}}(\Vec r)$ have interesting applications and for example, associated evolving ionisation fields can easily be generated from the sole knowledge of its power spectrum: one can imagine having a new class of fast forward modeling of the reionisation process within which we could vary astrophysical parameters (that are hopefully encoded in the reionisation times field power spectrum).
Conversely, the statistical properties of the aforementioned topological statistics could be used to constrain the power spectrum, under the GRF assumption, and by extension the physics that drives the propagation of radiation. Also, with a gaussian $t_{\text{reion}}(\Vec r)$, its power spectrum could be directly retrieved from measurements of its topological statistics. This could be interesting in the case where the power spectrum would be hardly measured. For instance, with the reconstructed reionisation times maps from 21 cm observations by the CNN mentioned before and in App. \ref{app:hiegel2022}, the power spectrum would be suppressed on small scales due the CNN smoothing and combining the diverse statistics measured in $t_{\text{reion}}(\Vec r)$ could help to obtain the proper power spectrum. Finally, by comparing the measurements and predictions, we show in this paper to what extent it is realistic to suppose that $t_{\text{reion}}(\Vec r)$ is a gaussian field, and we also infer a few causes of non-gaussianities.

\paragraph{Organisation of the paper}

We start by describing the $\emma$ and $\cmfast$ simulations we used in Sect. \ref{sec:simulations}, as well as some generated GRFs.
Afterwards follows a section gathering every topological characteristic we studied with the GRF theory, for which the behaviour is checked with the generated GRFs, as well as their measurements on the $\emma$ reionisation times fields (see Sect. \ref{sec:resultsEMMA}). 
The same analyses are also done for a $\cmfast$ simulation in Sect. \ref{sec:results21cmfast}.
Section \ref{sec:conclusion} presents our conclusions about this work and opens on a few perspectives. 
Appendix \ref{app:hiegel2022} presents some details about how we can reconstruct reionisation times maps from observation-like data.
Appendix \ref{app:std} details the calculation of the spectral moments from a specific power spectrum. Appendix \ref{app:PDFgradients} is the full calculation to obtain the PDF of the norm of the gradient of a GRF. Some results are also shown for the reionisation redshifts field in order to compare it to the reionisation times field in App. \ref{app:zreion}.
The cosmology parameters used are $(\Omega_m,\Omega_b,\Omega_{\Lambda},h,\sigma_8,n_s)=(0.31,0.05,0.69,0.68,0.81,0.97),$ as given by \citet{PlanckCollaboration2020}.

\paragraph{Notations}

Throughout this paper, we use the following notations that have been introduced in \citet{Pogosyan2009,Gay2012}. We call $F$ the studied field, which refers to the reionisation times fields. In this study, we work with normalised fields using the momenta of $F$ and its derivatives:
\begin{equation}
    \sigma_0^2 = \left< F^2\right>, \quad \sigma_1^2 = \left< \left( \Vec \nabla F\right)^2 \right>, \quad \text{and} \quad \sigma_2^2 = \left<\left(\Delta F \right)^2\right>. 
\label{eq:firstSigma}
\end{equation}

We introduce thereafter notations for the normalised fields and its derivatives:
\begin{equation}
    x = \frac{F}{\sigma_0}, \quad x_i = \frac{\nabla_i F}{\sigma_1},  \quad \text{and} \quad x_{ij} = \frac{\nabla_i\nabla_j F}{\sigma_2},
\label{eq:field_notations}
\end{equation}
with $i,j\in\{1,2\}$ for 2D fields. 
Here, we mainly work on the reionisation times field $t^*_{\text{reion}}(\Vec r)$ that we normalise as follows:
\begin{equation}
    x = t_{\text{reion}} = \frac{t_{\text{reion}}^*-\overline{t_{\text{reion}}^*}}{\sigma_0}, 
\end{equation}
with $\overline{t_{\text{reion}}^*}$ the mean of the field.
Besides, the normalised reionisation times fields are called $x$ in the following work. But, when we refer to the values of the field, we use the notation below:
\begin{equation}
    \nu = x(\Vec r).
\end{equation}
Since we work on normalised reionisation times fields, when $\nu<0$, we probe moments before the average reionisation times, and when $\nu>0$, we probe those after the average reionisation time. Also, low values of $\nu$ refers therefore to early reionisation times, and large values of $\nu$ to late reionisation times. 

We also introduce dimensionless spectral parameters:
\begin{equation}
    R_0 = \frac{\sigma_0}{\sigma_1}, \quad R_* = \frac{\sigma_1}{\sigma_2}, \quad \text{and} \quad \gamma = \frac{R_*}{R_0} = \frac{\sigma_1^2}{\sigma_0 \sigma_2}.
\label{eq:spectralParameters}
\end{equation}
These parameters can be analytically expressed if the power spectrum of the field is known. The calculation is shown in  App. \ref{app:std} for a specific type of power spectrum.

\section{Simulated data}
\label{sec:simulations}
\subsection{$\emma$ simulation}
\label{sec:EMMAsimu}

\begin{table}   
\centering                          
\begin{tabular}{c | c | c c c}       
\hline\hline    
     & $\overline{t_{\text{reion}}^*}$ & \multicolumn{3}{c}{$\sigma_0^{t_{\text{reion}}^*}$} \\
     $R_f$ & -- & 1 & 2 & 6 \\
\hline
    \texttt{EMMA [Myrs]} & 790 & 107 & 91.6 & 62.2 \\
    \texttt{21cmFAST [Myrs]} & 801 & 165 & 149 & 97.5 \\
\hline\hline    
     & $\overline{z_{\text{reion}}^*}$ & \multicolumn{3}{c}{$\sigma_0^{z_{\text{reion}}^*}$} \\
     $R_f$ & -- & 1 & 2 & 6 \\
\hline
    \texttt{EMMA} & 6.99 & 0.82 & 0.67 & 0.43 \\
    \texttt{21cmFAST} & 6.40 & 1.32 & 1.17 & 0.74 \\
\hline
\end{tabular}
\caption{Average and standard deviation of both $\emma$ and $\cmfast$ reionisation time/redshift fields. The standard deviations are computed thanks to the expression given in App. \ref{app:std}, and depend on the power spectrum parameters of the fields.}       
\label{table:Simu_avg_and_std}      
\end{table}

In this work, we used a $512^3$ cMpc$^3$ h$^{-3}$ cosmological simulation with a resolution of 1 cMpc$^3$ h$^{-3}$ detailed in \citet{Gillet2021}. It has been obtained with the cosmological code $\emma$ (Electromagnétisme et Mécanique sur Maille Adaptative, \citet{Aubert2015}), which is an adaptive mesh refinement (AMR) code that couples hydrodynamics and radiative transfer, and in which, light is described as a fluid (resolved using the moment-based M1 approximation, \citet{Aubert2006}). The $\emma$ simulation follows the cosmology given by \citet{PlanckCollaboration2020} and has no AMR, no reduced speed of light and a stellar particle mass of $10^8$ M$_\odot$. 

As this work is focused on 2D fields, a hundred of $512^2$ cMpc$^3$ h$^{-2}$ slices (spaced from each other by 5 slices) are extracted from the $t_{\text{reion}}(\Vec r)$ field. They are smoothed with a gaussian kernel of standard deviation $R_f \in \{1,2,6\}$ (see Sect. \ref{sec:smoothing}), and normalised as described in Sect. \ref{sec:introduction}. The average and standard deviation of the $\emma$ reionisation times field are given in Tab. \ref{table:Simu_avg_and_std}.

\subsection{$\cmfast$ simulation}
We compare the statistics measurements of the $\emma$ simulation to those of a semi-analytical simulation generated with $\cmfast$\footnote{https://github.com/andreimesinger/21cmFAST} (version 3.0.3; \citet{Mesinger2011,Muray2020}).
The size of the simulation box generated is $256^3$ cMpc$^3$ h$^{-3}$, again with a resolution of 1 cMpc$^3$ h$^{-3}$. The reionisation model used varies only two parameters from the default ones: the ionising efficiency of high redshift galaxies $\zeta=40$, and the virial temperature $T_{vir}=10^5$ K. $\zeta$ controls the number of photons emitted by galaxies: the higher it is, the faster the reionisation is. $T_{vir}$ is the minimum virial temperature allowing a halo to start forming stars. 
Those parameters are chosen to approximately match the reionisation history of the $\cmfast$ simulation with the one of the $\emma$ simulation. $\cmfast$ can provide us with the reionisation redshifts maps, that we can convert into reionisation times maps (with a given cosmology), similar to the $\emma$ ones. From this 3D simulation, 51 slices (of size $256^2$ cMpc$^2$ h$^{-2}$) can be extracted (again spaced from each other by 5 slices), and they are also smoothed and normalised the same way the $\emma$ slices are. The average and standard deviation of the $\cmfast$ reionisation times field are given in Tab. \ref{table:Simu_avg_and_std}.

\subsection{Choices for the simulation data sets}

\subsubsection{Reionisation time/redshift fields}

We also extracted the reionisation redshifts fields $z_{\text{reion}}(\Vec r)$ from both $\emma$ and $\cmfast$ simulations. The following statistics analyses are performed on both reionisation times and redshifts fields. In the main text, we only present the results of the $t_{\text{reion}}(\Vec r)$ field, and we briefly present a similar analysis of $z_{\text{reion}}(\Vec r)$ in App. \ref{app:zreion}.

\subsubsection{Smoothing}
\label{sec:smoothing}

\begin{table}   
\centering                          
\begin{tabular}{c | c c c | c}       
\hline\hline    
     & \multicolumn{3}{c}{Simulations} & SKA \\
\hline
    $R_f$ & 1 & 2 & 6 & -- \\    
    $\Delta_x$ [cMpc] & 1.48 & 2.96 & 8.88 & 8.3 \\  
    $\Delta_\theta$ [arcmin] & 0.57 & 1.14 & 3.42 & 3.11 \\
\hline
\end{tabular}
\caption{Angular resolutions corresponding to the size of the smoothing kernel applied to our simulation at a redshift $z=6.905$. The angular and spatial resolutions are also given for the radio-telescope SKA at the same redshift for a maximum baseline of 2 km \citep{Giri2018b}.}       
\label{table:angularResSKA}      
\end{table}

\begin{figure*}[ht]
   \centering
   \includegraphics[width=1\textwidth]{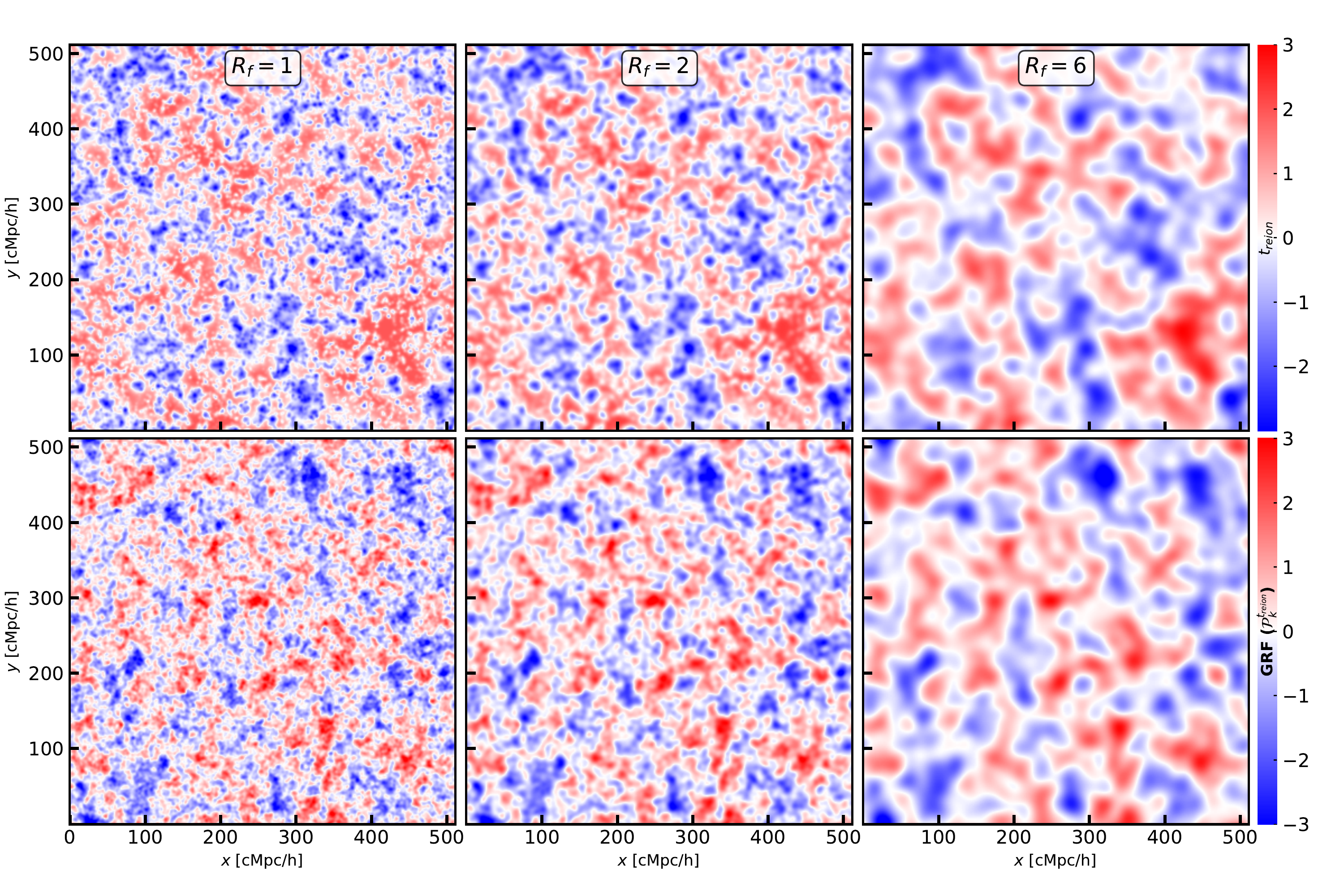}
    \caption{2D slices of the $\emma$ reionisation times field (first row) and the 2D GRFs obtained with the corresponding power spectrum (second row). Each column corresponds to different smoothings, with from left to right, $R_f\in\{1,2,6\}$. All the fields are normalised.}
    \label{fig:maps_EMMA_GRFs}
\end{figure*}

As mentioned above, we apply different smoothings on the reionisation times fields. We use gaussian kernels with different standard deviations. We chose to smooth the fields for the following reasons:
\begin{itemize}[topsep=0pt,itemsep=0pt]
    \item The future observed images from the SKA for example will have a lower spatial resolution than what we simulate here with $R_f\in\{1,2\}$ (see the angular resolutions given for the different kernel sizes in Tab. \ref{table:angularResSKA} compared to the equivalent values for SKA), which means that we will not probe the smallest scales of our simulations.
    \item In order to compute the field momenta or spectral parameters, we need to integrate over power spectra, which lead to possible divergence (see the shape of the spectra in section \ref{sec:GRFsimu}). Smoothing the fields avoid this divergence problem, and is also well accounted for in the GRF theory.
    \item We use the discrete persistent structure extractor ($\disperse$\footnote{http://www2.iap.fr/users/sousbie/web/html/indexd41d.html} ; \citet{Sousbie2011}), which assumes that the input field defines a Morse function, without large zero-gradient patches. As in \citet{Thelie2022}, we use it here to extract the critical points and the skeleton of $t_{\text{reion}}(\Vec r)$, and smoothing it prevents such patches to be present.
    \item The $\emma$ reionisation times fields are represented on the first row of Fig.\ref{fig:maps_EMMA_GRFs} for the different smoothings. The gaussian smoothing filters out the smallest structures, while keeping the global shape of the larger scales: we ‘gaussianise' the fields. Therefore, smoothing the reionisation times fields allows us to pinpoint the scales that are at the origin of non-gaussian features in our measures.
\end{itemize}

\subsection{Gaussian random fields (GRFs)}
\label{sec:GRFsimu}

A GRF is a gaussianly distributed field and if it has a null average, its PDF can be written as follows:
\begin{equation}
    P(\vec x) d^n\vec x = \frac{1}{(2\pi)^{n/2} \cdot \det(C)^{\frac{1}{2}}} \exp\left(-\frac{1}{2} \vec x \cdot C^{-1} \cdot \vec x\right) d^n\vec x,
\label{eq:PDF}
\end{equation}
where $\vec x$ is a n-D vector function of the position and $C = \left<\vec x \otimes \vec x\right>$ is the covariance matrix. For instance, $\vec x$ could be expressed as follows: $\vec x = (x, x_1, x_2,...)$ with the dimensionless field $x=F/\sigma_0$, and its first derivatives as defined in Eq. \ref{eq:field_notations}.

To generate and analytically study this kind of field, we only need its power spectrum as a GRF is entirely defined by it. From the power spectrum $\mathcal{P}_k$, we also have access to the momenta, as follows \citep{BBKS,Pogosyan2009,Gay2011}:
\begin{equation}
    \sigma_i^2 = \frac{ 2\pi^{\frac{d}{2}} }{ \Gamma\left(\frac{d}{2}\right) } \int_0^\infty k^{2i} \mathcal{P}_k k^{d-1} dk,
\end{equation}
where $i\in \mathbb{N}$ corresponds to the number of derivation of the field, and $d$ is the dimension of the field. The analytical derivation of the momenta is done in App. \ref{app:std} for a specific form of power spectrum detailed below.

\begin{figure}
   \centering
   \includegraphics[width=0.5\textwidth]{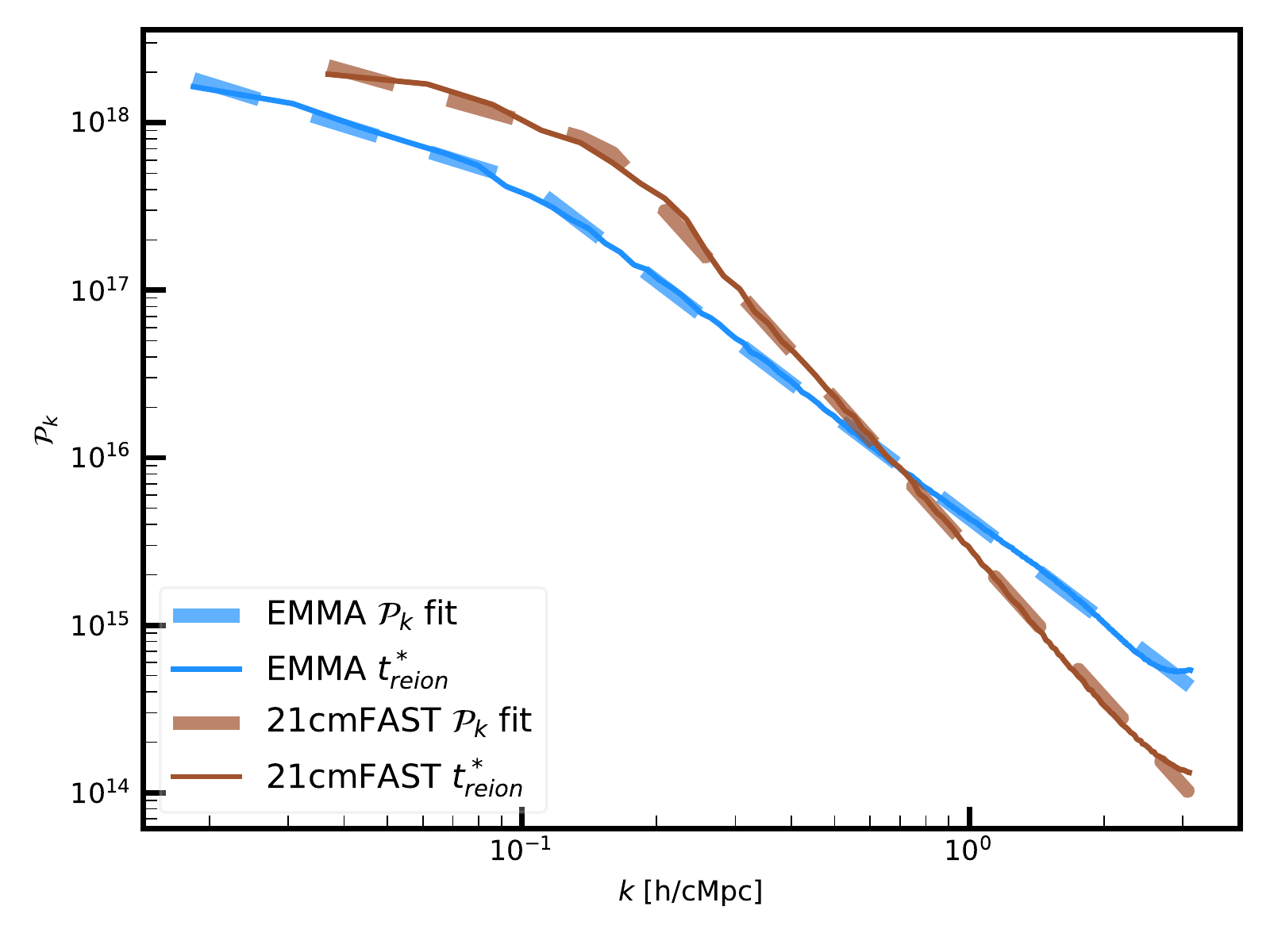}
    \caption{Fitting (in dashed lines) of the reionisation times power spectra. The straight lines corresponds to the power spectrum measured on the fields of each simulation ($\emma$ in blue and $\cmfast$ in brown). The fittings are done on the average logarithmic power spectrum of every 2D slices for each field. The fields are not normalised in order to do the fitting.}
    \label{fig:fit_Pk}
\end{figure}

We use the average power spectrum of the slices of our simulated fields: they are represented in logarithmic scales in Fig. \ref{fig:fit_Pk} for the $\emma$ and $\cmfast$ reionisation times fields. We use the following expression to fit these power spectra:
\begin{equation}
    \mathcal{P}_k = 
    \begin{cases}
        A_1 k^{n_1} \quad \text{if} \quad k\leq k_{\text{thresh}} \\
        A_2 k^{n_2} \quad \text{if} \quad k> k_{\text{thresh}} \\
    \end{cases}
    ,
\label{eq:Pk}
\end{equation}
with $A1$ and $A2$ the amplitude of each part and $n_1$ and $n_2$ the power of each part. $k_{\text{thresh}}$ is the threshold separating the two parts of the power spectrum. We obtain the parameters given in Table \ref{table:powerSpectraParams} after fitting the $\emma$ and $\cmfast$ reionisation times power spectra. In Fig. \ref{fig:fit_Pk}, the dashed lines represent the average expected power spectra in logarithmic scales of $t_{\text{reion}}(\Vec r)$ which fit pretty well the both simulations curves. This figure lets us also see that there are more large structure in the $\cmfast$ field than in the $\emma$ field, as well as less small structures.

\begin{table}   
\centering                          
\begin{tabular}{c | c c c c c c}       
\hline\hline                 
 & $A_1$ & $n_1$ & $A_2$ & $n_2$ & $k_{\text{thresh}}$ \\    
\hline
    & \multicolumn{5}{c}{\texttt{EMMA}} \\

    $t_{\text{reion}}$   & $6.70\times10^{16}$ & -0.83 & $4.34\times 10^{15}$ & -2.03 & 0.10 \\

    $z_{\text{reion}}$   & 4.03  & -0.75 & 0.38 & -1.80 & 0.10 \\
\hline
    & \multicolumn{5}{c}{\texttt{21cmFAST}} \\

    $t_{\text{reion}}$   & $1.83\times10^{17}$ & -0.75 & $2.87\times 10^{15}$ & -2.96 & 0.15 \\

    $z_{\text{reion}}$   & 6.17  & -0.91 & 0.25 & -2.86 & 0.20 \\
\end{tabular}
\caption{Parameters defining the power spectra of the reionisation times and redshifts fields ($t_{\text{reion}}(\Vec r)$ and $z_{\text{reion}}(\Vec r)$ for both $\emma$ and $\cmfast$ simulations. They are obtained thanks to a fitting of the power spectrum of the fields in order to generate GRFs in the same units as the reionisation times (years) or redshift before they are normalised.}   
\label{table:powerSpectraParams}      
\end{table}

The smoothed power spectrum of our GRF is defined as follows:
\begin{equation}
    \text{\large $\mathcal{P}_k^{\text{smoothed}} = $}
    \large
    \begin{cases}
        A_1 k^{n_1}\ e^{-2R_f^2k^2}\ /\ 2\pi \quad k\leq k_{\text{thresh}} \\
        A_2 k^{n_2}\ e^{-2R_f^2k^2}\ /\ 2\pi \quad k> k_{\text{thresh}} \\
    \end{cases},
\end{equation}
with $R_f$ the standard deviation of the kernel (i.e. size of the kernel), expressed in number of cells in the following analyses.

\begin{figure}
   \centering
   \includegraphics[width=0.5\textwidth]{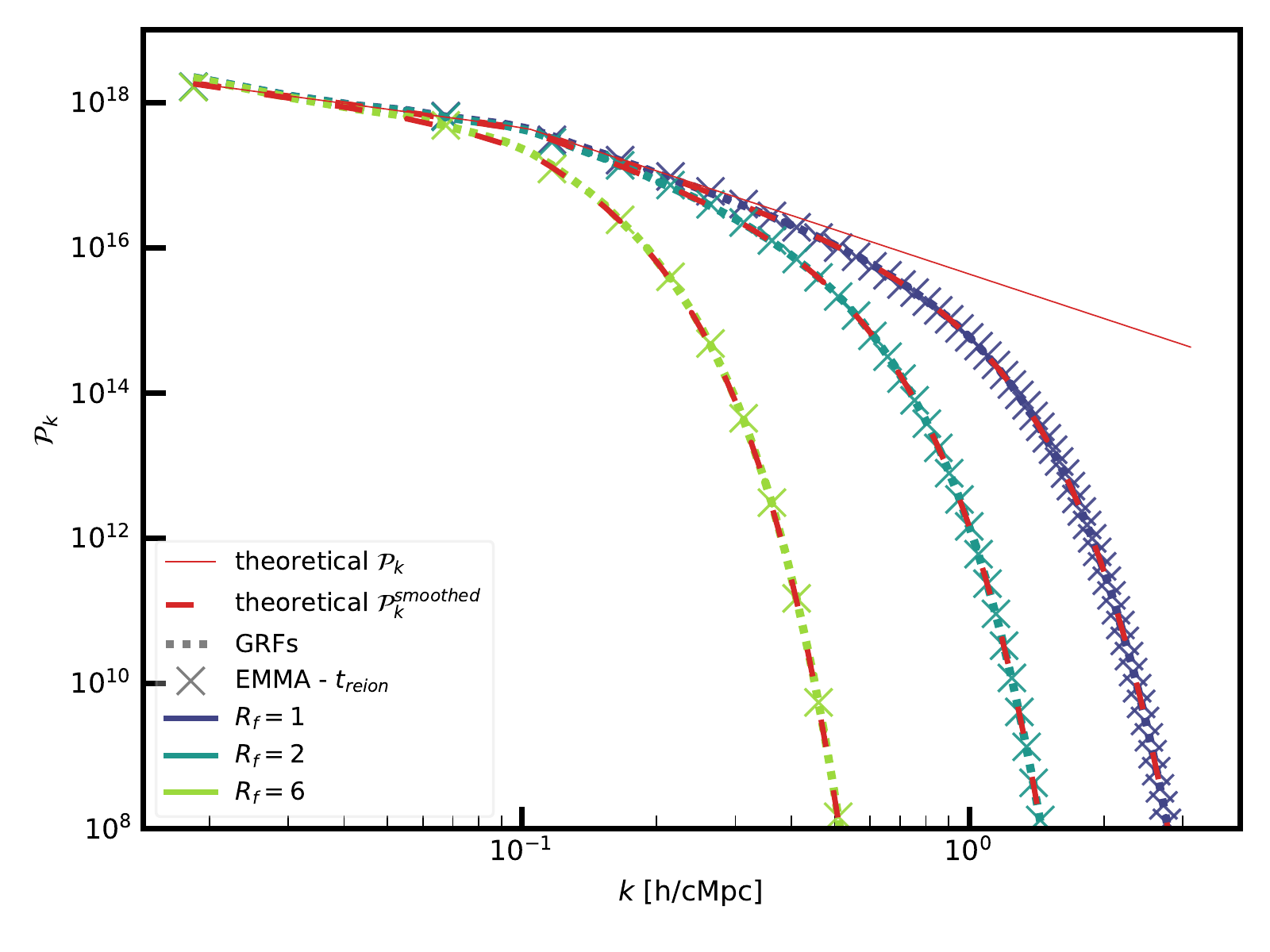}
    \caption{Power spectra of the GRFs (colored dotted lines) and $\emma$ reionisation times (colored crosses). All hundreds runs have been averaged to show a mean power spectrum for each set of simulation. The theoretical power spectra are shown with an incorporated smoothing and without it by the dashed and straight red lines respectively. Three different smoothings are represented with $R_f\in\{1,2,6\}$.}
    \label{fig:EMMA_GRFs_pk}
\end{figure}

We also generate multiple sets of a hundred of runs (with different seeds) of GRFs with the power spectrum of both $\emma$ and $\cmfast$ $t_{\text{reion}}(\Vec r)$ (for which, the parameters are given in the Table \ref{table:powerSpectraParams}). For each power spectrum, three sets of GRFs with different smoothing are created, with the following kernel sizes: $R_f \in \{1,2,6\}$. Besides, the GRFs are all normalised the same way as the simulations. Figure \ref{fig:maps_EMMA_GRFs} shows example of GRF maps (second row) with different smoothing (see each column): while differences can be spotted, such GRFs are close to the EMMA fields.  
Fig. \ref{fig:EMMA_GRFs_pk} shows the expected power spectra and the ones measured in the simulations: the GRF and $t_{\text{reion}}(\Vec r)$ curves are well superimposed. When we increase the kernel size $R_f$ (see the purple, blue and green coloured curves), larger and larger scales are smoothed, as expected. 

\section{Topological measurements on $\emma$ simulations and comparisons to GRF theory predictions}
\label{sec:resultsEMMA}

In this section, we extract several topological statistics from the $\emma$ reionisation times field $t_{\text{reion}}(\Vec r)$. We also derive their expression and compare them to the different runs of GRFs in order to statistically check the behaviour of our theoretical curves and our measures.

\subsection{Filling factor of the field: PDF and reionisation history}
\label{sec:filling_factor_emma}
\subsubsection{Measurements on the reionisation times field}

The filling factor of the reionisation times field $t_{\text{reion}}(\Vec r)$ shows the reionisation history (or fraction of ionised volume $Q_{\text{HII}}$) of the simulation box. It allows us to study the global evolution of the ionisation of the gas during the EoR. It can be directly extracted from the $t_{\text{reion}}(\Vec r)$ map by counting the number of values lower than a time threshold, which gives us the cumulated PDF of the $t_{\text{reion}}(\Vec r)$ values. The $t_{\text{reion}}(\Vec r)$ PDF tells us about the distribution of reionisation times in the box: it also incorporates reionisation evolution information. 
If both not-cumulated and cumulated distributions are symmetric with respect to the average reionisation time, then it means that the reionisation evolves the same way during all the EoR. If they are asymmetric and peaks at a larger time, then reionisation is slow before accelerating. On the contrary, if it peaks at a smaller time, then reionisation starts rapidly before slowing down.

\subsubsection{GRF theoretical expression}

We can rather directly compute the filling factor of a gaussian field with the gaussian field theory because it only requires the PDF of the value of the field \citep{Gay2012}:
\begin{equation}
    P(x) dx = \frac{1}{2\pi} e^{-\frac{1}{2}x^2} dx.
\label{eq:PDFFieldValues}
\end{equation}
To calculate the filling factor, the PDF has to only depend on the normalised field $x=F/\sigma_0$. Now, this statistic is the number of field values exceeding a given threshold $\nu$. Applied to our reionisation times fields, the number of values that has a higher time than a threshold is the same as the number of cells that are still neutral. It corresponds then directly to the fraction of neutral gas volume $Q_{\text{HI}}$.
However, in our case, we are interested in the fraction of ionised volume $Q_{\text{HII}} = 1-Q_{\text{HI}}$. It corresponds to the number of values smaller than a given threshold, as follows:
\begin{equation}
    Q_{\text{HII}}(\nu) = \int_0^\nu P(x)dx = \frac{1}{2} \erf\left(\frac{\nu}{\sqrt{2}}\right),
\label{eq:qhii_treion}
\end{equation}
where $\erf(\nu) = \frac{2}{\sqrt{\pi}} \int_0^\nu e^{-y^2}dy$.

\subsubsection{Comparison of the measurements and the predictions}

\begin{figure}
   \centering
   \includegraphics[width=0.5\textwidth]{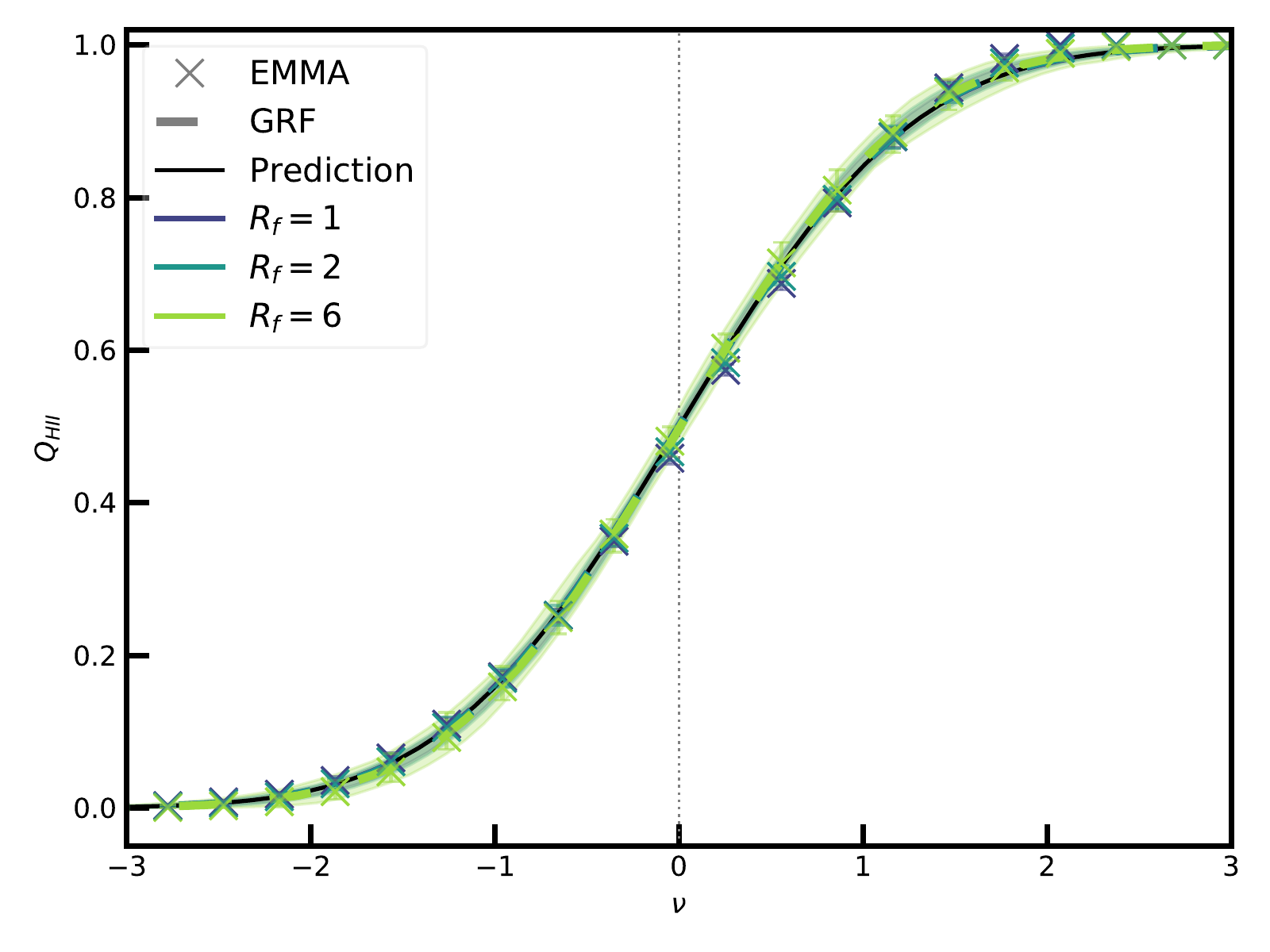}
    \caption{Fraction of ionised volume for the different smoothings (in colours). The median of every run is computed for each field. The dashed lines correspond to the GRFs, and the crosses are for the $\emma$ reionisation times field. The black lines are the theoretical predictions. The shaded areas and the error bars represent the dispersion around the median (1\textsuperscript{st} and 99\textsuperscript{th} percentiles) of the GRFs and $t_{\text{reion}}(\Vec r)$ respectively. Here, $\nu$ represents the value of the normalised reionisation times.}
    \label{fig:QHII_EMMA_GRFs_treion}
\end{figure}

\begin{figure}
   \centering
   \includegraphics[width=0.5\textwidth]{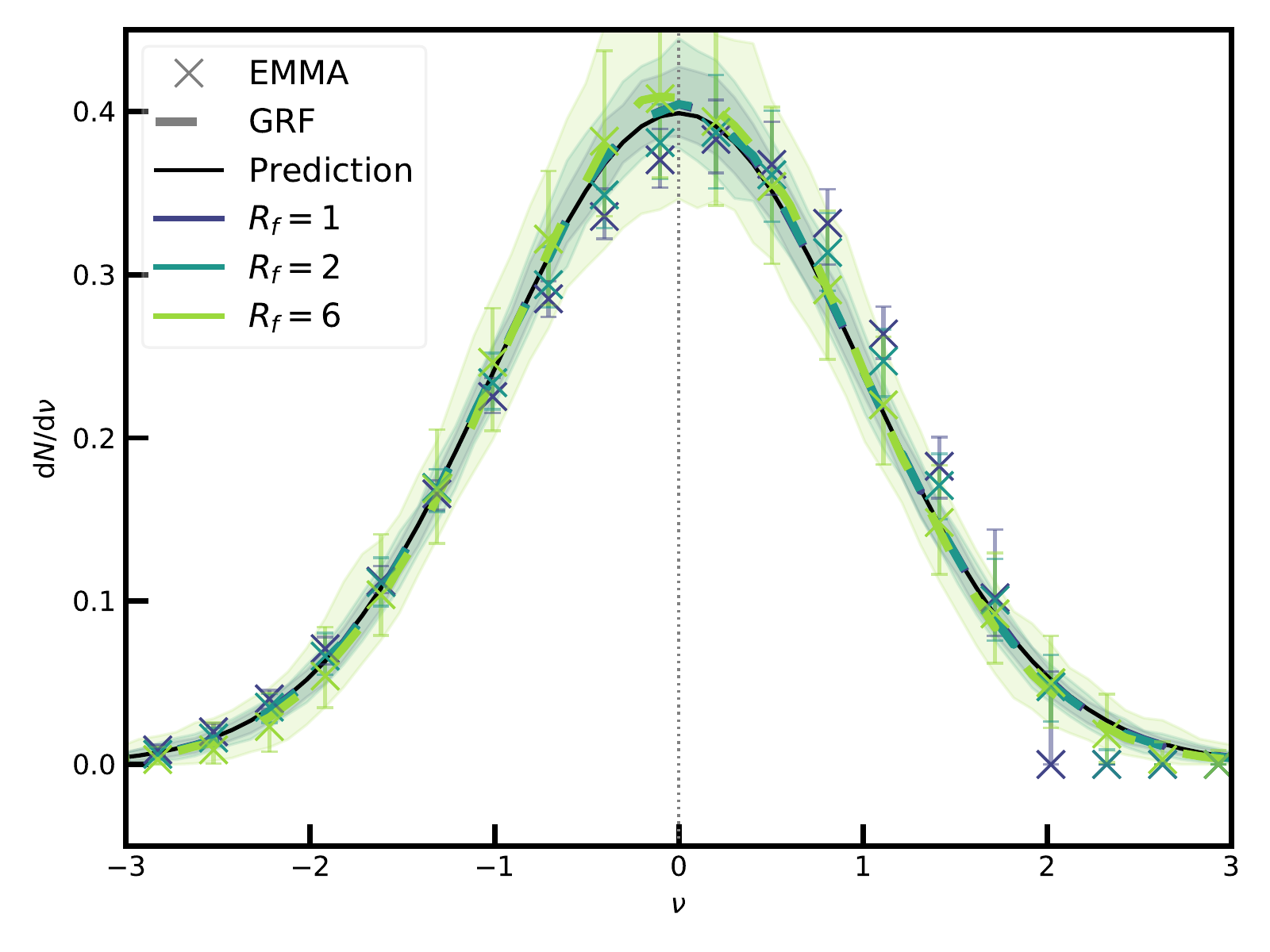}
    \caption{PDF of the median of every run of the fields for the different smoothings (in colours). The dashed lines correspond to the GRFs, and the crosses are for the $\emma$ fields. The black lines are the theoretical predictions. The shaded areas and the error bars represent the dispersion around the median (1\textsuperscript{st} and 99\textsuperscript{th} percentiles) of the GRFs and $t_{\text{reion}}(\Vec r)$ respectively. Here, $\nu$ represents the value of the normalised reionisation times.}
    \label{fig:QHII_EMMA_GRFs_nocumul_treion}
\end{figure}

We show the filling factors on Fig. \ref{fig:QHII_EMMA_GRFs_treion} and the PDFs on Fig. \ref{fig:QHII_EMMA_GRFs_nocumul_treion}: the crosses (and error bars) are the $\emma$ measures, the dashed lines (and the shaded areas) are the GRFs measures, and the GRF predictions are shown in black.
Firstly, in both figures, the GRF distributions follow well the predictions by the GRF theory. 
The filling factor measurements on the $\emma$ reionisation times fields are rather close to the predictions, depicting a rather symmetric reionisation process.
In Fig. \ref{fig:QHII_EMMA_GRFs_nocumul_treion}, the PDF measured on the $\emma$ $t_{\text{reion}}(\Vec r)$ maps are however not symmetric, the peak being shifted toward later times: the filling factors or cumulated PDFs hide this imprint of non-gaussianity. At the same time, when we smooth the fields on larger areas (i.e. $R_f$ increases), the distributions tend to become more symmetric around the mean reionisation time $\nu=0$ (i.e. close to the GRF predictions). 

Therefore, the regions that ionise after the average time of the simulation ($\nu=0$, or around 790 Myrs or a redshift of 7, see Tab. \ref{table:Simu_avg_and_std}) cause the asymmetry. It means that the reionisation process is a little bit slower at early times and accelerates afterwards. \citet{Mellema2006} or \citet{Dixon2016} have shown with the brightness temperature field that the asymmetry arises toward the end of the EoR. This asymmetry is also a key parameter in the newly developed code $\amber$ \citep{Trac2021}, in which we can directly tune an asymmetry parameter of the reionisation history. 
We see here that the non-gaussianity of the reionisation fields are filtered out with the smoothing: they are thus hidden in the time differences on small scales structures, and at later times.

\subsection{PDF of the gradient norm field: ionising front velocities}
\subsubsection{Measurements on the reionisation times field}

\begin{figure*}[ht]
   \centering
   \includegraphics[width=1\textwidth]{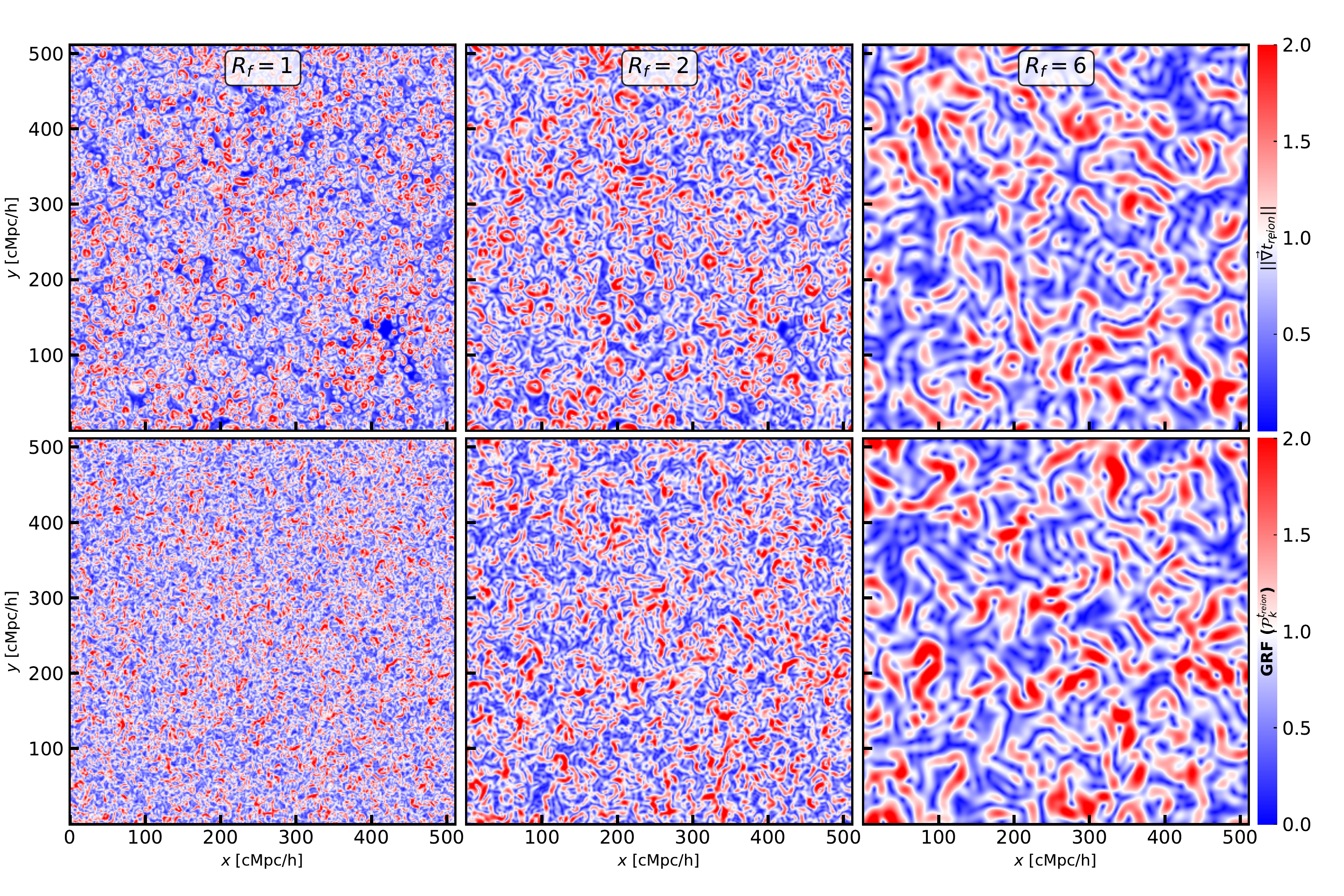}
    \caption{2D slices of the norm of the $\emma$ reionisation times gradient (first row) and the norm of the GRF gradient obtained with the corresponding power spectrum (second row). Each column corresponds to different smoothings, with from left to right, $R_f\in\{1,2,6\}$.}
    \label{fig:maps_grad_EMMA_GRFs}
\end{figure*}

In this section, we analyse the norm of the spatial gradients of each field that we define as follows, for a GRF or a reionisation times field $F$:
\begin{equation}
    \| \vec \nabla x \| = \frac{1}{\sigma_1} \sqrt{\left(\nabla_1 F\right)^2+\left(\nabla_2 F\right)^2} = R_0 \sqrt{\left(\nabla_1 x\right)^2+\left(\nabla_2 x\right)^2},
\end{equation}
with $\nabla_i x$ for $i\in\{1,2\}$ are the two components of the gradient of the field $x=F/\sigma_0$, and $R_0$ is given in Eq. \ref{eq:spectralParameters}.
Numerically, the gradient of the fields are computed thanks to Fourier transforms. Each component of the gradient is obtained as follows: 
\begin{equation}
    \nabla_i x = \mathcal{F}^{-1}[\Tilde{x} \times ik_i],
\end{equation}
where $\Tilde{x} = \mathcal{F}[x]$ is the Fourier transform of the field, and $k^2 = k_1^2 + k_2^2$. We should note that here we observe in 2D a 3D phenomenon, which means that the gradient norms is probably underestimated as we miss the third direction component of the front velocities. Figure \ref{fig:maps_grad_EMMA_GRFs} shows gradient norm maps of the reionisation times fields on the first row, as well as those of the GRFs on the second row for each smoothing ($R_f\in\{1,2,6\}$, see the columns). The maps are rather close to the GRF ones by eye for $R_f\in\{2,6\}$. However, some disparities start to be visible for the smallest smoothing kernel $R_f=1$ again: the reionisation times field has larger structures than the corresponding GRF. 

The gradients norm of $t_{\text{treion}}(\Vec r)$ is linked to the reionisation velocity field defined by \citet{Deparis2019}, which is the inverse of the spatial derivative of the reionisation times field. It contains information about the ionising front velocity, and \citet{Deparis2019} have shown that these fronts move forward in two stages. They are first slowed down by dense neutral gas and their speed is smaller than the speed of light. However, when reaching the end of the EoR, the fronts accelerate because radiations reach underdense regions. It means that as time increases, the reionisation front speeds increase, or conversely, the gradient norms of reionisation times field decrease.

\subsubsection{GRF theoretical expression}

The PDF of the gradient norm of a gaussian field $\| \vec\nabla F \|$ only depends on the field ($x=F/\sigma_0$) and its first derivatives ($x_1$ and $x_2$ as defined in Eq. \ref{eq:field_notations}), such as:
\begin{equation}
    P(x,x_1,x_2) dxdx_1dx_2 = \frac{2}{(2\pi)^{3/2}} e^{-\left(\frac{1}{2} x^2 + x_1^2 + x_2^2\right)} dxdx_1dx_2.
\label{eq:PDFgradnorm}
\end{equation}
This joint PDF is obtained quite easily with Eq. \ref{eq:PDF} with a 3-dimensional covariance matrix (as shown in App. \ref{app:PDFgradients}). From this expression, thanks to an integral over the field values and a change of variable, we can retrieve the PDF of the norm of the field gradient:
\begin{equation}
    2\pi P(w)wdw = 2w e^{-w^2} dw \quad \text{with} \quad w^2=x_1^2+x_2^2.
\end{equation}

\subsubsection{Comparison of the measurements and the predictions}

\begin{figure*}
   \centering
   \includegraphics[width=1\textwidth]{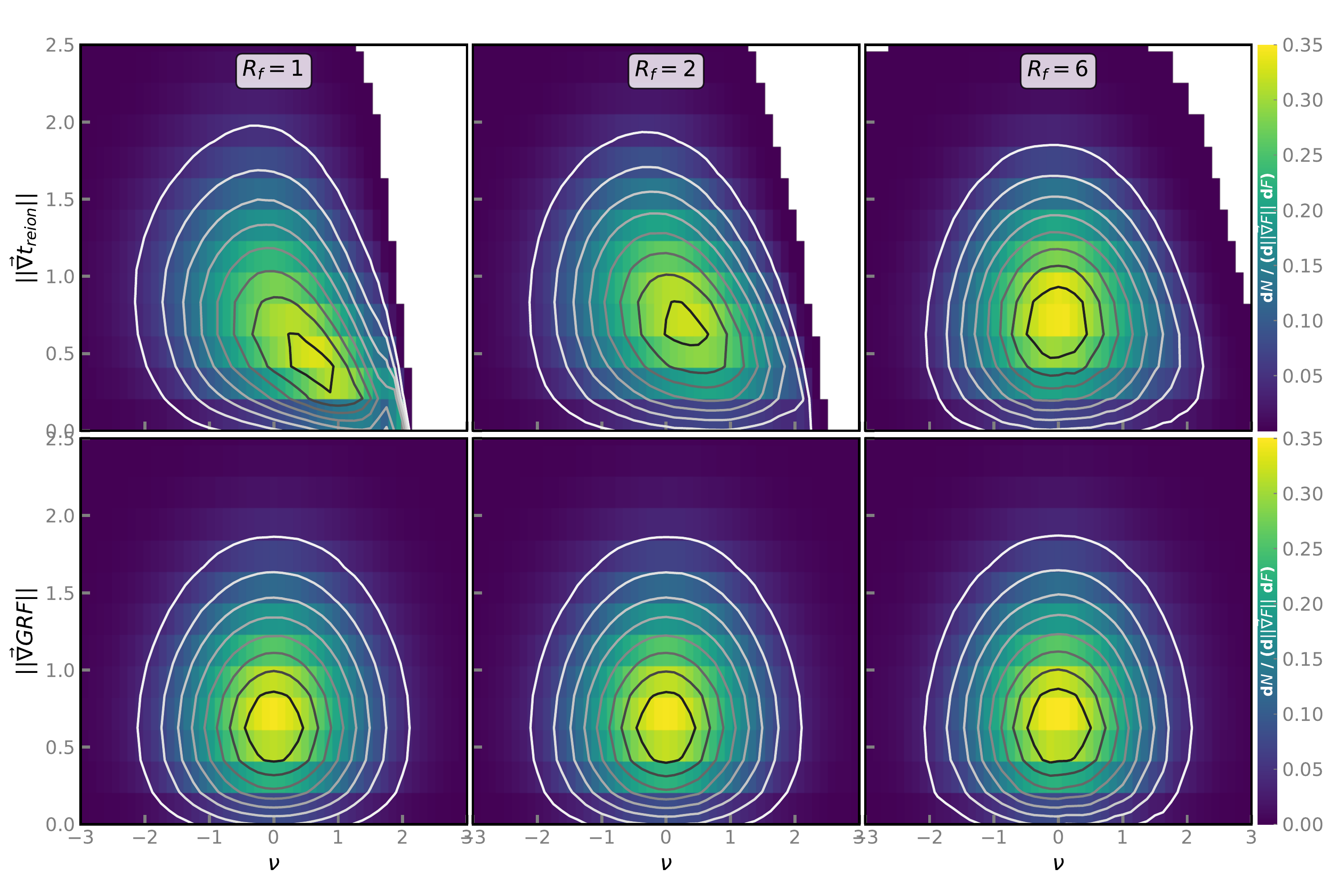}
    \caption{2D PDFs of the gradient norms with respect to the values of the fields of every run for each field and different smoothings ($R_f\in\{1,2,6\}$, see each column). The first row corresponds to the $\emma$ reionisation times, and the second row to their corresponding GRFs. The gray-scale lines are the isocontours of the histograms. Here, $\nu$ represents the value of the normalised reionisation times.}
    \label{fig:Map_histgrad_EMMA_GRFs_treion}
\end{figure*}

\begin{figure}
   \centering
   \includegraphics[width=0.5\textwidth]{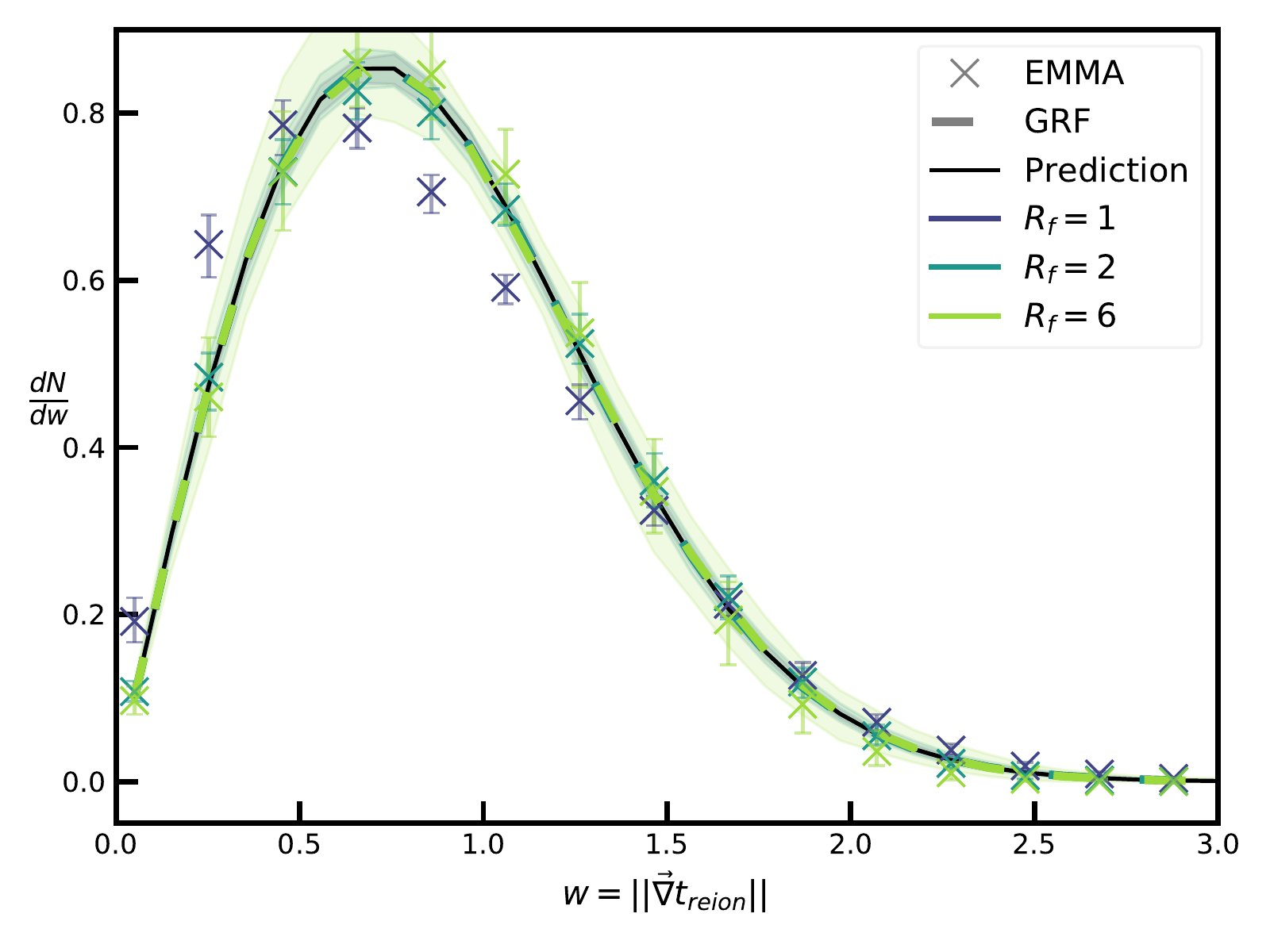}
    \caption{Filling factors of the gradient norms of the fields for the different smoothings (in colours). The median of every run is computed for each field. The dashed lines correspond to the GRFs, and the crosses are for the $\emma$ reionisation times fields. The black lines are the theoretical predictions. The shaded areas and the error bars represent the dispersion around the median (1\textsuperscript{st} and 99\textsuperscript{th} percentiles) of the GRFs and $t_{\text{reion}}(\Vec r)$ respectively. Here, $w$ represents the value of the gradients norm of the reionisation times.}
    \label{fig:normGrad_EMMA_GRFs_treion}
\end{figure}

First, we have an insight on the radiation fronts velocities at each time of the EoR with 2D distributions of the gradients norm of $t_{\text{reion}}(\Vec r)$ with respect to the field values. Figure \ref{fig:Map_histgrad_EMMA_GRFs_treion} shows these 2D PDFs for the $\emma$ reionisation times field on first row and their corresponding GRFs on second row. Each smoothing kernel size are represented with $R_f\in\{1,2,6\}$ in the columns. The GRFs cases are symmetric around their mean reionisation time $\nu=0$, as expected. We can see again the asymmetry mentioned earlier for the $\emma$ reionisation times fields: the peak is shifted towards the larger times. This asymmetry enables us to see the acceleration of the ionising fronts as the EoR progress.
If we integrate along the y-axis, we retrieve the PDF of the gradients norm, which is shown in Fig. \ref{fig:normGrad_EMMA_GRFs_treion}. The dashed lines represent the GRFs measurements, the GRF prediction is shown with the black line, and the crosses are for the $\emma$ reionisation times fields. The GRFs measurements are superimposed to the predictions, and the $\emma$ measurements underestimate a little bit the gradients norm, due to the acceleration of the radiation fronts at the end of the EoR.
On both 2D and 1D distributions, the GRFs measurements are independent on the smoothing, as expected since Eq. \ref{eq:PDFgradnorm} do not depend on the kernel size. For the $\emma$ measurements, increasing the kernel size makes them closer to the predictions.

Our measurements on the $\emma$ reionisation times field reflect thus the increase in the ionising front velocities as time increases, and they are probably a strong cause of the asymmetry of our reionisation times field distributions, and therefore of the non-gaussianity of the process. Besides, this phenomenon mainly impacts the small scales of $t_{\text{reion}}(\Vec r)$ as it tends to be filtered out with large smoothing.

\subsection{Isocontours length: size evolution of ionised and neutral bubbles}
\subsubsection{Measurements on the reionisation times field}

\begin{figure*}[ht]
   \centering
   \includegraphics[width=1\textwidth]{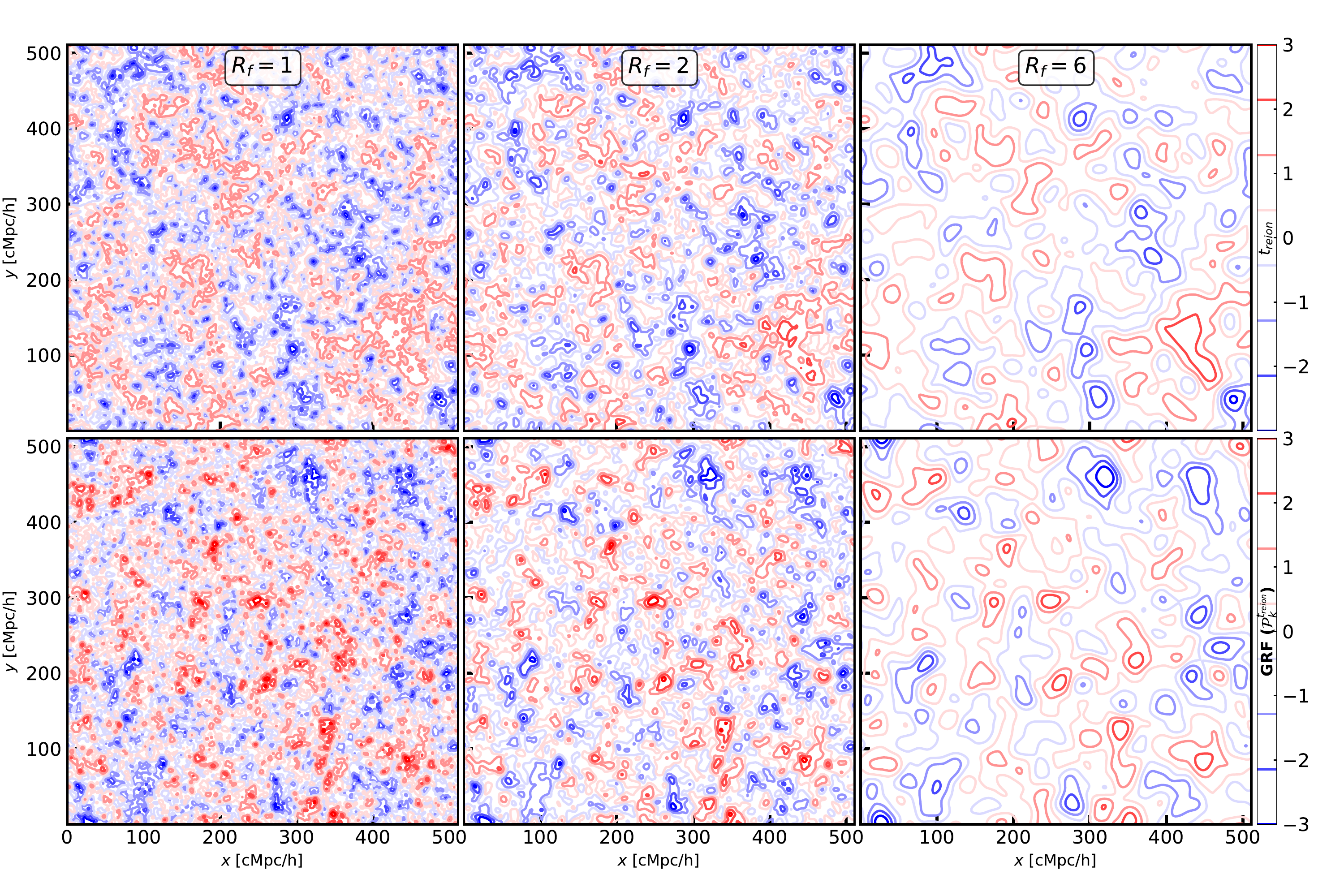}
    \caption{Isocontours of 2D slices of the $\emma$ reionisation times field (first row) and of the 2D GRF obtained with the corresponding power spectrum (second row). Each column corresponds to different smoothings, with from left to right, $R_f\in\{1,2,6\}$. All the fields are normalised. Eight levels of contours are represented with the colors.}
    \label{fig:maps_contours_EMMA_GRFs}
\end{figure*}

The isocontours of the reionisation times field allow us to know how far radiation propagates at a specific time. Figure \ref{fig:maps_contours_EMMA_GRFs} shows the isocontours of the $\emma$ reionisation times field on the first row and of its corresponding GRFs that uses the same power spectrum on the second row for the three smoothings (see the three columns). The bluest contours represent the earliest times, and the reddest ones represent the latest times. The number of contours per level visually decreases the larger the gaussian kernel is (because small structures disappear when increasing $R_f$).

The isocontours length $\mathcal{L}$ of $t_{\text{reion}}(\Vec r)$ informs us about the extent of a reionisation time level. On Fig. \ref{fig:maps_contours_EMMA_GRFs}, we can see that as the reionisation times increases, the isocontours encompass first larger and larger regions (blue contours), and after the mean reionisation time $\nu=0$, smaller and smaller regions (red contours). Their length contains therefore information on the size of the ionised/neutral bubbles, on the percolation of the ionised bubbles and the different reionisation stages.

\begin{figure}
   \centering
   \includegraphics[width=0.5\textwidth]{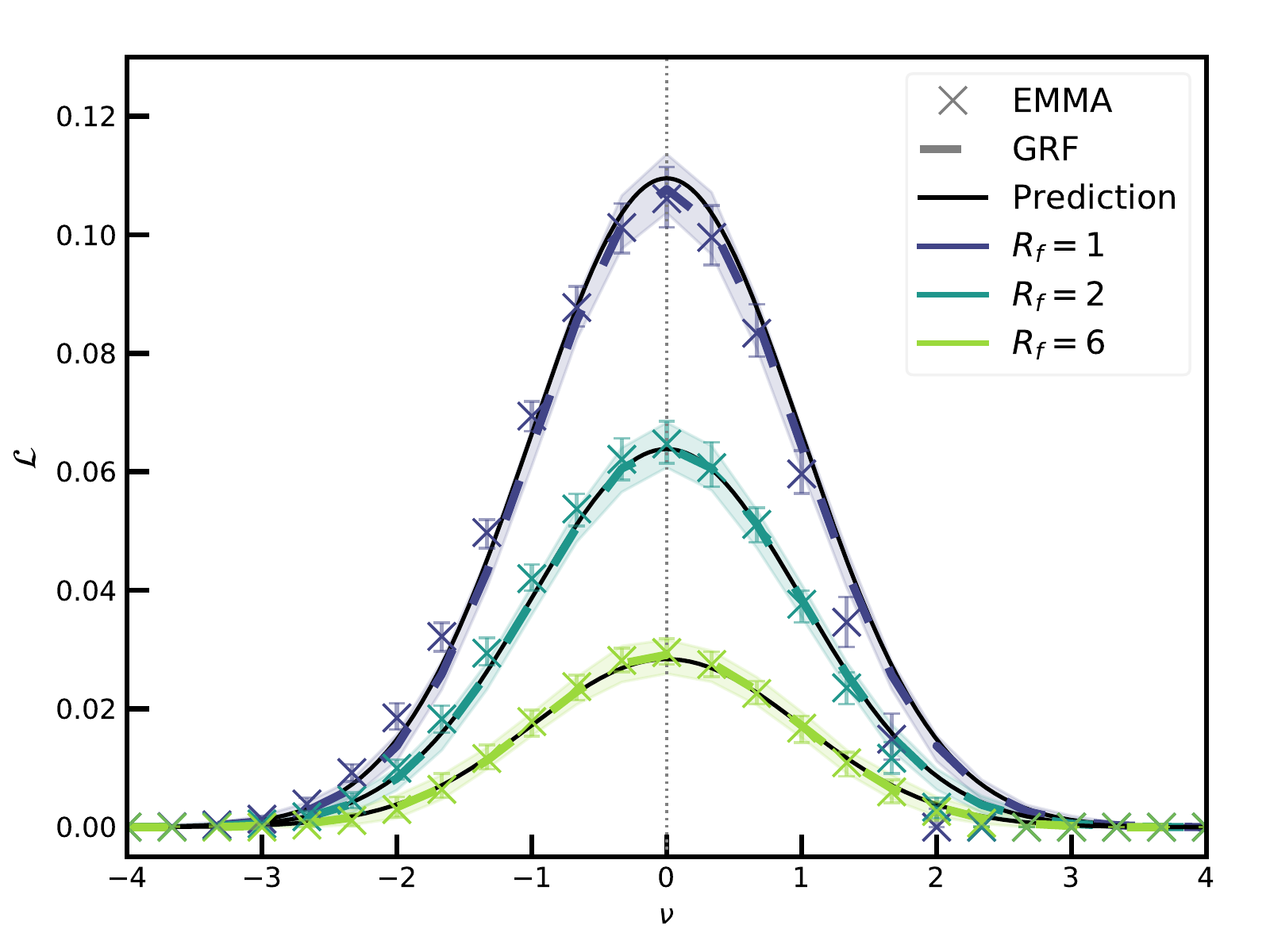}
    \caption{Distribution of the isocontour length for the different smoothings (in colours). The median of every run is computed for each field. The dashed lines correspond to the GRFs, and the crosses are for the $\emma$ reionisation times field. The black lines are the theoretical predictions. The shaded areas and the error bars represent the dispersion around the median (1\textsuperscript{st} and 99\textsuperscript{th} percentiles) of the GRFs and $t_{\text{reion}}(\Vec r)$ respectively. Here, $\nu$ represents the value of the normalised reionisation times.}
    \label{fig:distrib_Lcont_EMMA_GRFs_treion}
\end{figure}

\subsubsection{GRF theoretical expression}

The average 2D isocontours length allows us to characterise the levels of a field by a measurement of their length. It is in fact one of the Minkowski functionals \citep{Schmalzing1998,Matsubara2003}, which have often been used to quantify the topology of the EoR (see e.g. \citet{Gleser2006,Lee2008,Friedrich2011,Hong2014,Yoshiura2017,Chen2019,Pathak2022}). The isocontours length at a level $\nu$ can be defined as follows \citep{Schmalzing1998,Matsubara2003,Gay2012}:
\begin{equation}
    \mathcal{L}(\nu) = \left< \frac{1}{R_0} \delta(x-\nu) (x_1^2+x_2^2)^{1/2}  \right> = \left< \frac{1}{R_0} \delta(x-\nu) w \right>,
\label{eq:Lcont_def}
\end{equation}
where $\delta$ is the Dirac delta distribution. $x$, $x_1$, and $x_2$ are the normalised field and first derivatives as defined in Eq. \ref{eq:field_notations}, and $w^2=x_1^2+x_2^2$. $R_0$ is defined in Eq. \ref{eq:spectralParameters} and corresponds to the ratio of the two first spectral momenta. It appears due to the normalisation of the field and its derivatives.

As $\mathcal{L}$ depends both on the field and its first derivative,  we need to use $P(x,x_1,x_2)$, as defined in Eq. \ref{eq:PDFgradnorm}, in order to calculate it. This probability has the following expression:
\begin{equation}
    2\pi P(x,w) wdxdw = \frac{2}{\sqrt{2\pi}} e^{-\left(\frac{1}{2} x^2 + w^2\right)} wdxdw.
\end{equation}
Now, we compute the isocontour length by integrating the quantity of Eq. \ref{eq:Lcont_def} over the field and gradients norm values:
\begin{equation}
    \mathcal{L}(\nu) = \int_{-\infty}^\infty dx \int_0^\infty dw\ \frac{2\pi}{R_0} P(x,w) \delta(x-\nu) w^2 = \frac{1}{2\sqrt{2}R_0} e^{-\frac{1}{2}\nu^2}.
\end{equation}

\subsubsection{Comparison of the measurements and the predictions}

The isocontours length $\mathcal{L(\nu)}$ distributions are shown on Fig. \ref{fig:distrib_Lcont_EMMA_GRFs_treion} with dashed lines for the GRFs, black lines for the GRF prediction, and crosses for the $\emma$ reionisation times field $t_{\text{reion}}(\Vec r)$ for the three smoothings. The measurements of the GRFs distributions follow well the GRF prediction curves, as well as the $\emma$ measurements. The contours grow in length until the mean reionisation time $\nu=0$ and then their length decreases. The measurements and predictions vary with the smoothing: the larger the gaussian kernel is, the smaller the isocontours length is. Indeed, with small $R_f$, there are much more isocontours per reionisation level (i.e. more small structures), making a total length of isocontours larger than for large $R_f$.
 
The evolution of the isocontours length with time traces as expected the evolution of the ionised bubbles and neutral regions. At the beginning of the EoR, the first ionised bubbles appear because the gas starts to be ionised. It corresponds to the dark blue contours of the bottom left panel of Fig. \ref{fig:maps_contours_EMMA_GRFs}. As reionisation progresses, the bubbles grow, traced by the larger contours represented by light blue levels in the same figure. The contours increase in length and sometimes two small dark blue contours merge into one, allowing us to have an insight on the percolation process. At $\nu\sim0$, the isocontours have reached a maximal length with the ionised regions intertwined with the neutral regions. Afterwards, the isocontours length $\mathcal{L}$ decreases while the isocontours start to encompass the last neutral regions, until $\mathcal{L}$ is near 0 at the moment when there is almost no more neutral gas.
The process we mention here is very close to what \citet{Chen2019} have found with several phases during the reionisation process, starting with an ionised bubble stage, followed by an ionised fibre stage, when the bubbles have merged into long fibres throughout the box. Then, there is the sponge stage: it is the moment when the ionised fibres are intertwined with the neutral fibres. The process ends by a neutral fibre stage and a neutral bubble stage.

With this statistics, the $\emma$ reionisation times do not show much imprints of non-gaussianities as it follows the GRF predictions, and more so for the largest smoothing. 

\subsection{PDF of minima values: reionisation seeds counts}
\label{sec:ReionisationSeedCount}
\subsubsection{Measurements on the reionisation times field}

In order to compare our simulations to the predicted reionisation seed count, we use the topological code $\disperse$ \citep{Sousbie2011}, that allows us to extract the critical points of a field. $\disperse$ relies on the Morse theory to get topological information from the fields by the study of differentiable functions.
It is ran on every 2D slice of reionisation times fields, as well as on every GRF generated, with a $10^{-5} - \sigma$ persistence threshold\footnote{The persistence is very low here so that we apply no selection on the extracted critical points. It is a threshold controlling the maximal distance between the field values in a maximum/minimum pair in the extracted features by $\disperse$. It allows to controls the significance of the extracted topological features, and therefore the smoothness of the features. It can be used to override the noise of the input field. More details are given by \cite{Sousbie2011} or \citet{Thelie2022}.}.

We focus on the minima of the reionisation times field. We call them the ‘seeds', or equivalently the ‘sources', of reionisation because they are the regions where the gas is locally reionised the first. We can measure the number of minima at a reionisation time (i.e. the PDF of the $t_{\text{reion}}(\Vec r)$ values at its minima). Counting these reionisation seeds informs us about the evolution of the EoR: for example if the PDF peaks at early times, it means that the majority of reionisation seeds appear at the beginning of the EoR, whereas if the PDF is uniformly distributed, then sources contribute to reionisation during all the EoR in an equivalent manner.

\subsubsection{GRF theoretical expression}
\label{sec:distrib_cp}

The PDF of the minima of the reionisation times field can also be derived with the GRF theory. To compute it, we need a joint PDF of the field that is dependent on the field, and on its first and second derivatives. Indeed, the critical points corresponds to the zeros of the first derivative, and the sign of the second derivative gives us an information on the type of critical point (it is positive for minima). In that case and for a GRF, the calculation is in 6 dimensions in 2D and requires changes of variables to make the covariance matrix diagonal: 
\begin{equation}
    u = -(x_{11} + x_{22}), \quad v = \frac{1}{2} (x_{11}-x_{22}), \quad \zeta = \frac{x-\gamma u}{\sqrt{1-\gamma^2}},
\end{equation}
where $\gamma = \sigma_1^2/(\sigma_0\sigma_2)$. $x$, $x_{11}$, and $x_{22}$ corresponds to the dimensionless field and derivatives as defined in Eq. \ref{eq:field_notations}. The full calculation is done in \citet{Gay2011} and the resulting PDF is the following:
\begin{equation}
    P(\zeta,x_1,x_2,u,v,x_{12}) = \frac{16}{(2\pi)^3} e^{-Q(\zeta,x_1,x_2,u,v,x_{12})},
\end{equation}
with
\begin{equation}
    Q(\zeta,x_1,x_2,u,v,x_{12}) = \frac{1}{2}\zeta^2 + x_1^2 + x_2^2 + \frac{1}{2}u^2 + 4v^2 + 4x_{12}^2,
\end{equation}
with $x_1$, $x_2$, and $x_{12}$ dimensionless derivatives of the field as defined in Eq. \ref{eq:field_notations}.

The average extrema density is given by the following expression (which is explained in \citet{Gay2011}):
\begin{equation}
    \frac{\partial n_{\text{ext}}}{\partial \nu} = \left< \frac{1}{R_*^2} \left|x_{11}x_{22} - x_{12}^2\right| \delta(x_{1}) \delta(x_{2}) \delta(x-\nu) \right>.
\label{eq:MinDist_avgExp}
\end{equation}
The non trivial part of the minima distribution calculation is the 6D integration involved in Eq. \ref{eq:MinDist_avgExp}. A version of this integration for 3D fields is detailed in App. A of \citet{BBKS}. Additional constraints (on the eigenvalues of the hessian matrix of the field) are required to make Eq. \ref{eq:MinDist_avgExp} a distribution of minima. Here, we only give the resulting distribution of minima:
\begin{equation}
\footnotesize
\begin{split}
    \frac{\partial n_{\text{min}}}{\partial \nu}   & = \frac{ \exp\left({\frac{-\nu^2}{2}}\right) }{ \sqrt{2\pi}R_*^2 } \left[1 - \erf{ \left(\frac{\gamma \nu}{\sqrt{2(1-\gamma^2}} \right) }\right] K_1(\nu,\gamma) \\
    & + \frac{ \exp\left(\frac{-3\nu^2}{6-4\gamma^2}\right) }{ \sqrt{2\pi(1-\frac{2}{3}\gamma^2)}R_*^2 } \left[1 - \erf{ \left(\frac{\gamma \nu}{\sqrt{2(1-\gamma^2)(3-2\gamma^2)}} \right)} \right] K_2 \\
    & - \frac{ \exp\left(\frac{-\nu^2}{2(1-\gamma^2)}\right) }{ \sqrt{2\pi(1-\gamma^2)}R_*^2 } \left[1 + \exp\left( \frac{\nu^2}{2(1-\gamma^2)} \right) \right] K_3(\nu,\gamma), \\
\end{split}
\label{eq:dn_dnu}
\end{equation}
with $R_*$ defined in Eq. \ref{eq:spectralParameters}, and
\begin{equation}
\text{\footnotesize $
    K_1(\nu,\gamma) = \frac{\gamma^2 (\nu^2-1)}{8\pi}, \quad
    K_2 = \frac{1}{8\pi\sqrt{3}}, \quad \text{and} \quad
    K_3(\nu,\gamma) = \frac{\gamma(1-\gamma^2) \nu}{2(2\pi)^{3/2}}. 
$}
\end{equation}

\subsubsection{Comparison of the measurements and the predictions}

\begin{figure}
   \centering
   \includegraphics[width=0.5\textwidth]{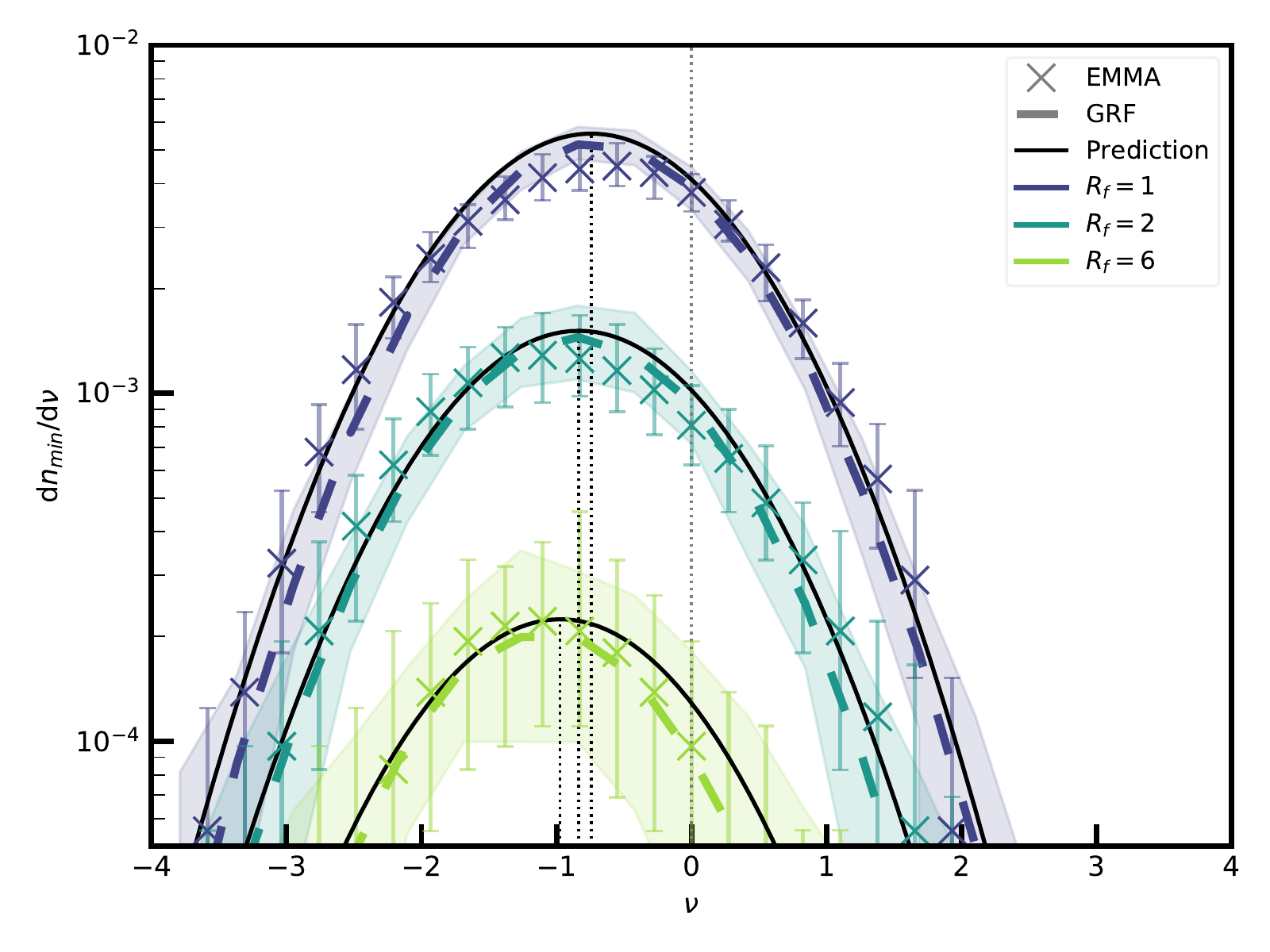}
    \caption{Distribution of the critical points of the fields for the different smoothings (in colours). The median of every run is computed for each field. The dashed lines correspond to the GRFs, and the crosses are for the $\emma$ reionisation times fields. The black lines are the theoretical predictions. The shaded areas and the error bars represent the dispersion around the median (1\textsuperscript{st} and 99\textsuperscript{th} percentiles) of the GRFs and $t_{\text{reion}}(\Vec r)$ respectively. The black dotted vertical lines represent the average of the predictions. Here, $\nu$ represents the value of the normalised reionisation times.}
    \label{fig:distrib_cp_EMMA_GRFs_treion}
\end{figure}

The PDFs of the minima values are shown on Fig. \ref{fig:distrib_cp_EMMA_GRFs_treion} with dashed lines for the GRFs, full black lines for the predictions, and crosses for the $\emma$ reionisation times field. The three smoothings are represented (see the three $R_f$ values in colours). 
The theoretical PDFs are centered around early reionisation times (i.e. $\nu<0$) because we look for minima. The GRFs and $\emma$ $t_{\text{reion}}(\Vec r)$ fields follow well the expected curves. When the fields are smoothed with a larger kernel (increasing $R_f$), the less significant minima get smoothed out, decreasing the number of counted minima. We can also note that smoothing on larger areas causes the distributions to be shifted towards smaller values of the field (i.e. towards the beginning of the EoR). 

With these results, we can thus see that when the EoR begins (smallest $\nu$), there are a few reionisation seeds that represent the first radiation sources of the Universe. The further the reionisation progresses, the more the number of reionisation seeds increases as more and more galaxies are created. It lasts until some intermediate time, where the minima distributions peaks, which happens before the average reionisation time at the mean reionisation time $\nu=0$. Then, the distributions decrease because more and more intergalactic gas is reionised by the already existing radiation sources, until it is all reionised. At this time, no more new seeds of ionised fronts propagation appear, even if there can be new sources in already reionised places.
The impact of the smoothing is that only the most exceptional and early reionisation seeds, which are significant enough, remain in the smoothed fields, while the other ones are filtered out. It means that these exceptional seeds reionise the other ones in an inside-out way.
Globally, there is little to no imprints of non-gaussianities with this statistic within the error bars.

\subsection{Skeleton length: regions of ionising fronts percolation}
\label{sec:skeletonSec}
\subsubsection{Measurements on the reionisation times field}

Thanks to $\disperse$, we can also extract the skeleton of $t_{\text{reion}}(\Vec r)$ from every reionisation times fields and GRFs, using the same persistence value as in Sect. \ref{sec:ReionisationSeedCount}. The skeleton is, as explained in Sect. \ref{sec:introduction}, the network formed by all the segments joining the maxima and saddle points together along the ridge of the field. In 2D, the skeleton corresponds to the edges of the reionisation patches, which are regions under the radiative influence of a reionisation seed. These edges, or the place where the patches intersect, are also the front lines between the radiation fronts of reionisation.
Looking at the skeleton (in black) on the top right panel of Fig. \ref{fig:Map_EMMA_treion_project}, we can see that it seems to have preferential directions along the diagonal of the map. They are due to diamond shapes produced around sources that are caused by the M1 approximation used in $\emma$ to model the radiative transfer \citep{Aubert2006}.

We are interested in the length distribution of the skeleton as a function of the time. As an example, if the distributions are as narrow as a Dirac distribution, then it means that the reionisation seeds are uniformly distributed in space so that all the radiation fronts encounter opposite fronts all at the same time. If the Dirac distribution peaks at $\nu<0$, then the percolation happens at early times, and if it peaks at $\nu>0$, then the percolation happens at late times. 
Otherwise, if the distribution is wider, it means that the seeds are not uniformly distributed and that the merger of the reionisation patches happens throughout the reionisation process.
If there were many ionised bubbles at early times and then a single growing bubble, then the distribution would be asymmetric. 
As the skeleton is the place where radiation fronts meet, it is hence impacted by the front velocities: if the fronts propagate faster, they can reach more distant regions, meaning that the reionisation patches are larger. If ones assumes a simple 2D lattice of circular patches of radius $R$, it is expected to have a total skeleton length per unit surface $L \sim 2\pi R/\pi R^2 \sim R^{-1}$ : scenario with large patches should lead to small $L$ values. Also, the accelerated fronts can cause the percolation to happen more rapidly in the remaining neutral regions, causing the distribution to decrease more sharply at the end of the EoR, and to be therefore asymmetric with respect to time.
Generally speaking, these distributions tell us if the ionising fronts percolate in a longer or shorter period of time. 

\subsubsection{GRF theoretical expression}

The analytical calculation of the skeleton length distribution is described in a few articles, such as \citet{Pogosyan2009,Gay2011,Gay2012}. The calculation are done in detail by \citet{Pogosyan2009} for example. 
In summary, the skeleton corresponds to all the points for which the gradient is aligned with an eigenvector of the hessian $\mathcal{H} = \Vec \nabla \Vec \nabla F$ of the field, following the gradient in the direction of the positive eigenvalue of the hessian.\footnote{To come back to our representation of the topology through a mountain landscape, the skeleton, and therefore the ridge lines are topologically called a critical lines. Walking on a ridge line is coming from a pass and going in the direction that goes up until the peak: it means that we follow the gradient in the direction of the positive eigenvalue of the hessian. Besides, as a comment, if we were to follow the gradient in the direction of the negative eigenvalue of the hessian, we would be on the anti-skeleton, aiming at reaching a minima.}
This definition can be mathematically written as follows: $\mathcal{H}\cdot \Vec \nabla F = \lambda \Vec \nabla F$ with $\lambda$ the positive eigenvalue of the hessian. In an equivalent way, a point is on a critical line if $s = \det(\mathcal{H}\cdot \Vec \nabla x, \lambda \Vec \nabla x)=0$, with the dimensionless field $x$ ($=F/\sigma_0$). 
The skeleton length distribution can therefore be written as follows:
\begin{equation}
    \frac{\partial L^{\text{skel}}}{\partial \nu} = \left< \frac{1}{R_*} \delta(s) |\Vec \nabla s| \delta(x-\nu) \right>,
\label{eq:Lskel_avgExp}
\end{equation}
with $|\Vec \nabla s|$ giving the length of the critical lines, and $R_*$ given in Eq. \ref{eq:spectralParameters}. The integration of Eq. \ref{eq:Lskel_avgExp} involves the PDF of the field, and its first, second and third derivative. The ‘stiff approximation' allows us to neglect the derivatives of order higher than 3, which simplify the calculations. The critical lines are therefore considered rather straight: the total length of skeleton may therefore probably be reduced with this approximation. 
The final expression of the skeleton length per unit total field surface:
\begin{equation}
\footnotesize
\begin{split}
    \frac{\partial L^\text{skel}}{\partial \nu} & = \frac{1}{\sqrt{2\pi} R_*} \exp^{-\frac{\nu^2}{2}} \Biggr[ \frac{1}{8\sqrt{\pi}} (\sqrt{\pi} + 2 \gamma \nu) \left( 1 + \erf\left( \frac{\gamma \nu}{\sqrt{2(1-\gamma^2)}} \right) \right) \\
    & +  \frac{\sqrt{1-\gamma^2}}{2\sqrt{2}\pi} \exp\left( - \frac{\gamma^2\nu^2}{2(1-\gamma^2} \right) \Biggr], \\
\end{split}
\label{eq:Lskel}
\end{equation}
with $\gamma$ defined in Eq. \ref{eq:spectralParameters}.

The total skeleton length of a field can be obtained by integrating the distribution of Eq. \ref{eq:Lskel} over all values of $\nu$. The resulting expression is the following \citep{Pogosyan2009,Gay2011}:
\begin{equation}
    L^\text{tot} = \left( \frac{1}{8} + \frac{\sqrt{2}}{4\pi} \right) \frac{1}{R_*}.
\label{eq:total_Lskel}
\end{equation}

\subsubsection{Comparison of the measurements and the predictions}

\begin{figure}
   \centering
   \includegraphics[width=0.5\textwidth]{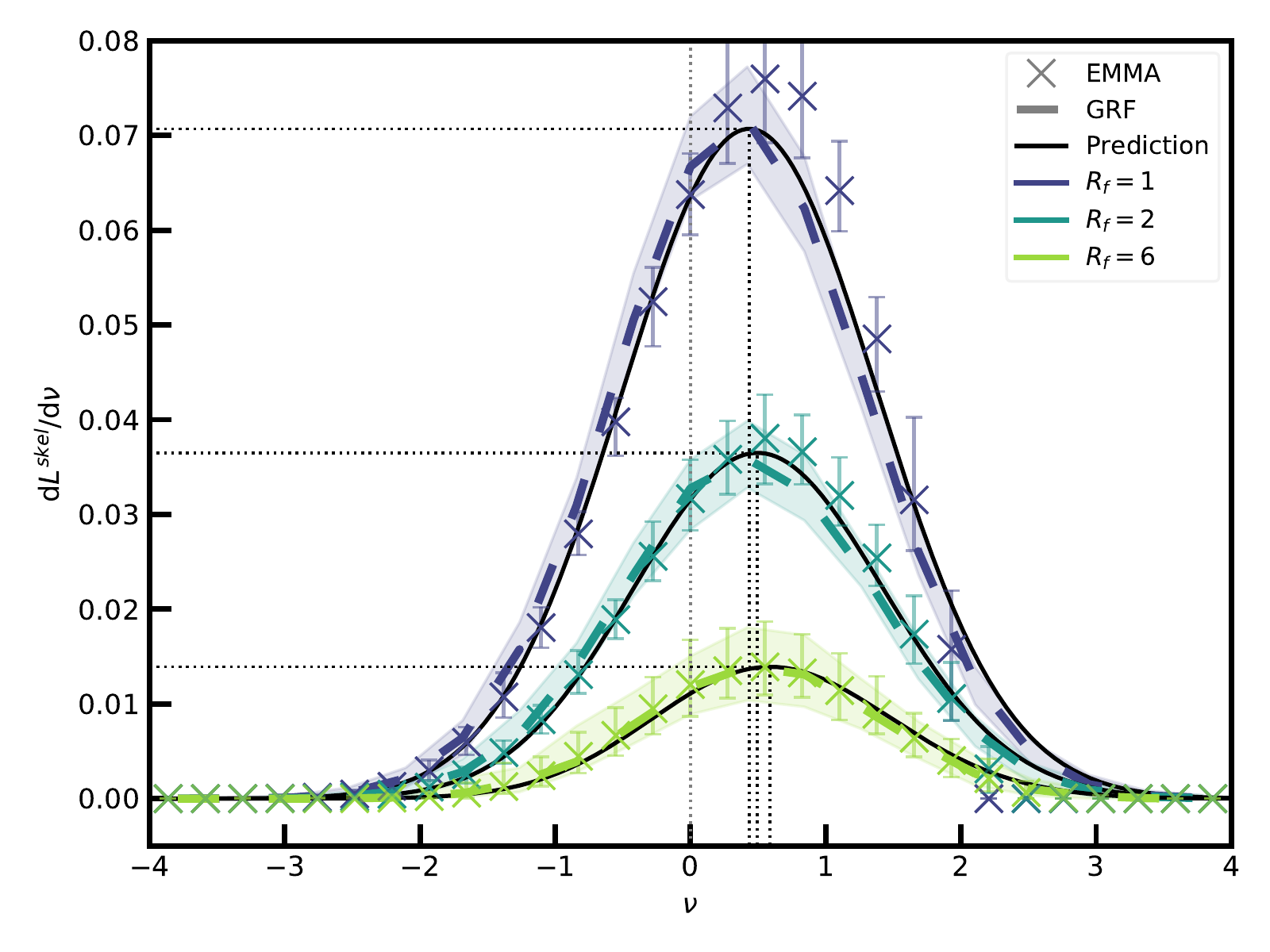}
    \caption{Distribution of the skeleton length of the fields for the different smoothings (in colours). The median of every run is computed for each field. The dashed lines correspond to the GRFs, and the crosses are for the $\emma$ reionisation times fields. The black lines are the theoretical predictions. The shaded areas and the error bars represent the dispersion around the median (1\textsuperscript{st} and 99\textsuperscript{th} percentiles) of the GRFs and $t_{\text{reion}}(\Vec r)$ respectively. The black dotted vertical lines represent the average of the predictions. Here, $\nu$ represents the value of the normalised reionisation times.}
    \label{fig:Lskel_EMMA_GRFs_treion}
\end{figure}

The skeleton length PDFs are shown in Fig. \ref{fig:Lskel_EMMA_GRFs_treion} with dashed lines for the GRFs, full black lines for the predictions, and crosses for the $\emma$ reionisation times field. The colours represent the three different smoothings (see the three $R_f$ values in colours). 
The ‘stiff approximation' lead to discrepancies between the measurements and the predictions. Besides, the analytical calculation of the skeleton length is local and $\disperse$ gives a global skeleton \citep{Pogosyan2009,Gay2012}, meaning that the predictions tend to underestimate the skeleton length by probably missing filaments. For these reasons, we multiply our measurements by a normalisation factor\footnote{The normalisation factor is given by the ratio between the total length of the measured skeleton over the one of the prediction. It depends on the smoothing via the spectral parameter $R_*$ (see Eq. \ref{eq:total_Lskel}).} to match the predictions amplitude following \citet{Gay2012}. Still, the shapes of the distribution seem to be conserved as the re-normalised measurements on the GRFs are almost superimposed to the predicted curves. The measurements on the reionisation times fields from $\emma$ are also pretty close to the predictions, especially when increasing $R_f$. The PDFs increase until reaching a maximum after the average reionisation time ($\nu>0$), before decreasing again. They are not centered on the average reionisation time $\nu=0$, shifted to the largest times when the smoothings increase, and asymmetric, except for the highest smoothings ($R_f=6$), for which the measured $\emma$ distributions are ‘symmetrised' thanks to the smoothing, and therefore closer to the predictions.

With this statistic, we can again follow the evolution of the EoR from the point of view of merging radiation fronts. During the beginning of the EoR, the ionised bubbles grow, and the radiation fronts start to reach farther out, increasing the skeleton length. At a reionisation time $\nu>0$, the distribution peaks when the ionising fronts percolate on the longer length: at this moment there are large ionised fibres. Afterwards, most gas is reionised, and there are less percolation of ionised bubbles, meaning that less ionised fronts encounter other fronts, until the gas is totally ionised at the end of the EoR ($\nu\sim3$ as in the reionisation history). 
With small smoothing kernel sizes ($R_f\in\{1,2\}$), we retrieve, as expected, the asymmetry that results from the increase of the radiation front velocities at the end of the EoR.

\section{Predictions from gaussian random field theories compared to $\cmfast$ simulations}
\label{sec:results21cmfast}

In this section, we compare the theoretical statistics described in Sect. \ref{sec:resultsEMMA} to the $\cmfast$ reionisation times field. For the sake of brevity, we only show the 2D histograms of the field values and its gradient norms, the fraction of ionised volume of gas, as well as the skeleton length distribution for the $\cmfast$ reionisation times fields and the GRFs generated with the corresponding power spectrum. Indeed, the $\cmfast$ reionisation fields give similar results to the $\emma$ ones: the semi-analytical generated field is generally close to the GRF predictions. It depicts an almost gaussian behaviour in every statistic and smoothing studied in this paper, as for the $\emma$ simulation, albeit a little bit less gaussian for the minima and skeleton length distributions. Again, increasing the gaussian kernel size (i.e. increasing $R_f$) tends to ‘gaussianise' the reionisation times field.

\begin{figure*}
   \centering
   \includegraphics[width=1\textwidth]{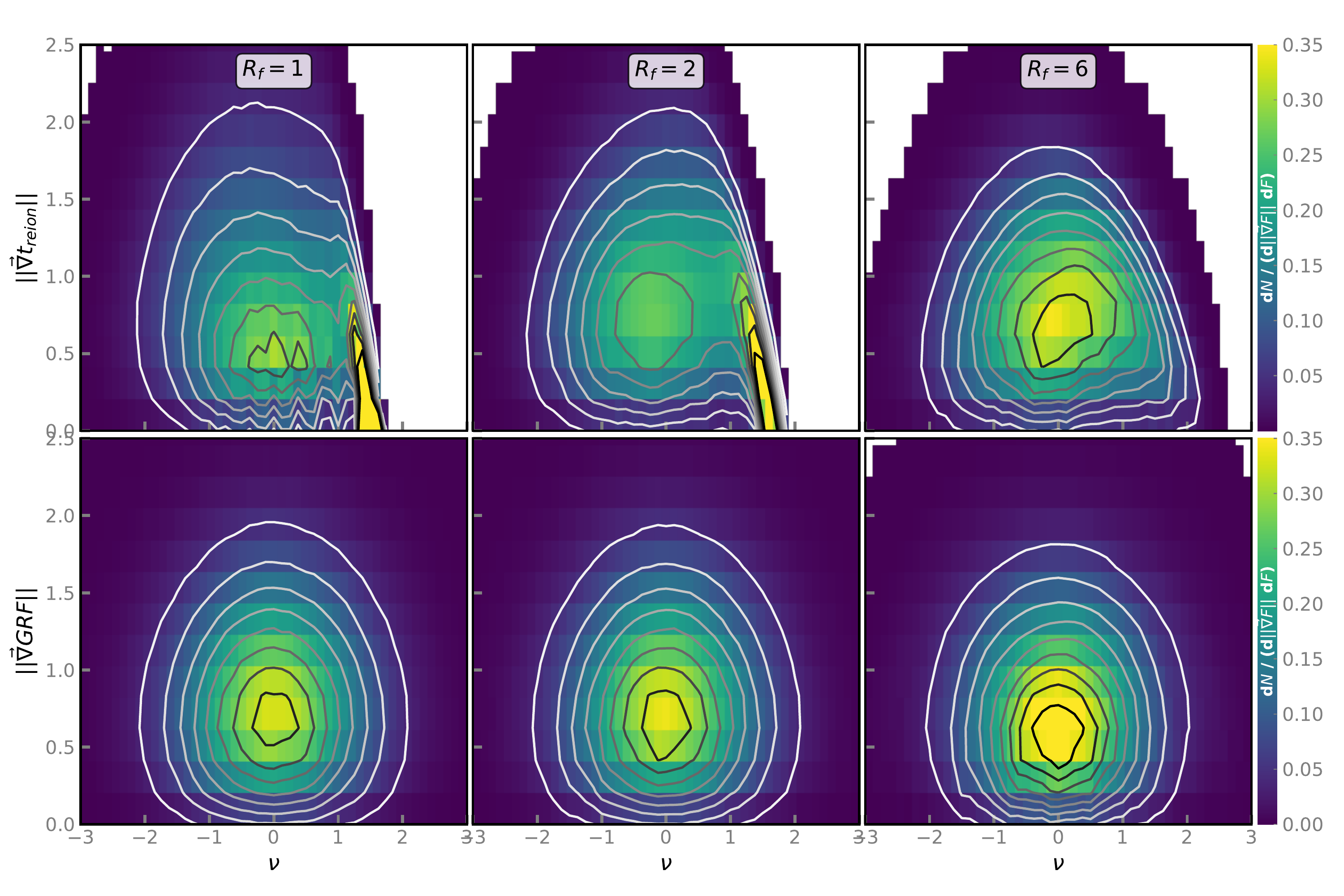}
    \caption{2D PDFs of the gradient norms with respect to the values of the fields of every run for each field and different smoothings ($R_f\in\{1,2,6\}$, see each column). The first row corresponds to the $\cmfast$ reionisation times, and the second row to their corresponding GRFs. The gray-scale lines are the isocontours of the histograms. Here, $\nu$ represents the value of the normalised reionisation times.}
    \label{fig:Map_histgrad_21cmFAST_GRFs_treion}
\end{figure*}

\begin{figure}
  \centering
  \includegraphics[width=0.5\textwidth]{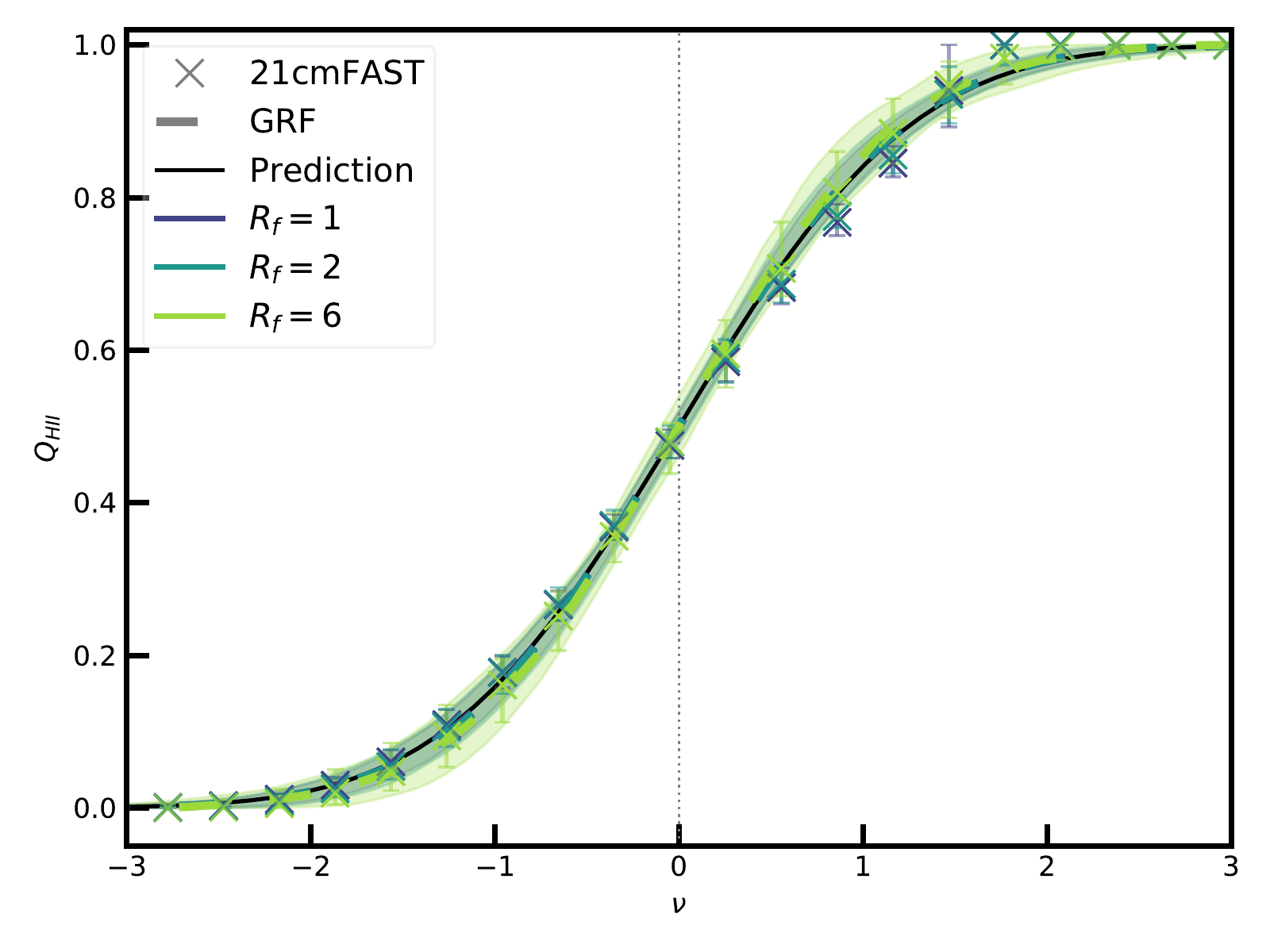}
    \caption{Fraction of ionised volume of the fields for the different smoothings (in colours). The median of every run is computed for each field. The dashed lines correspond to the GRFs, and the crosses are for the $\cmfast$ reionisation times fields. The black lines are the theoretical predictions. The shaded areas and the error bars represent the dispersion around the median (1\textsuperscript{st} and 99\textsuperscript{th} percentiles) of the GRFs and $t_{\text{reion}}(\Vec r)$ respectively. Here, $\nu$ represents the value of the normalised reionisation times.}
    \label{fig:QHII_21cmFAST_GRFs_treion}
\end{figure}

\begin{figure}
   \centering
   \includegraphics[width=0.5\textwidth]{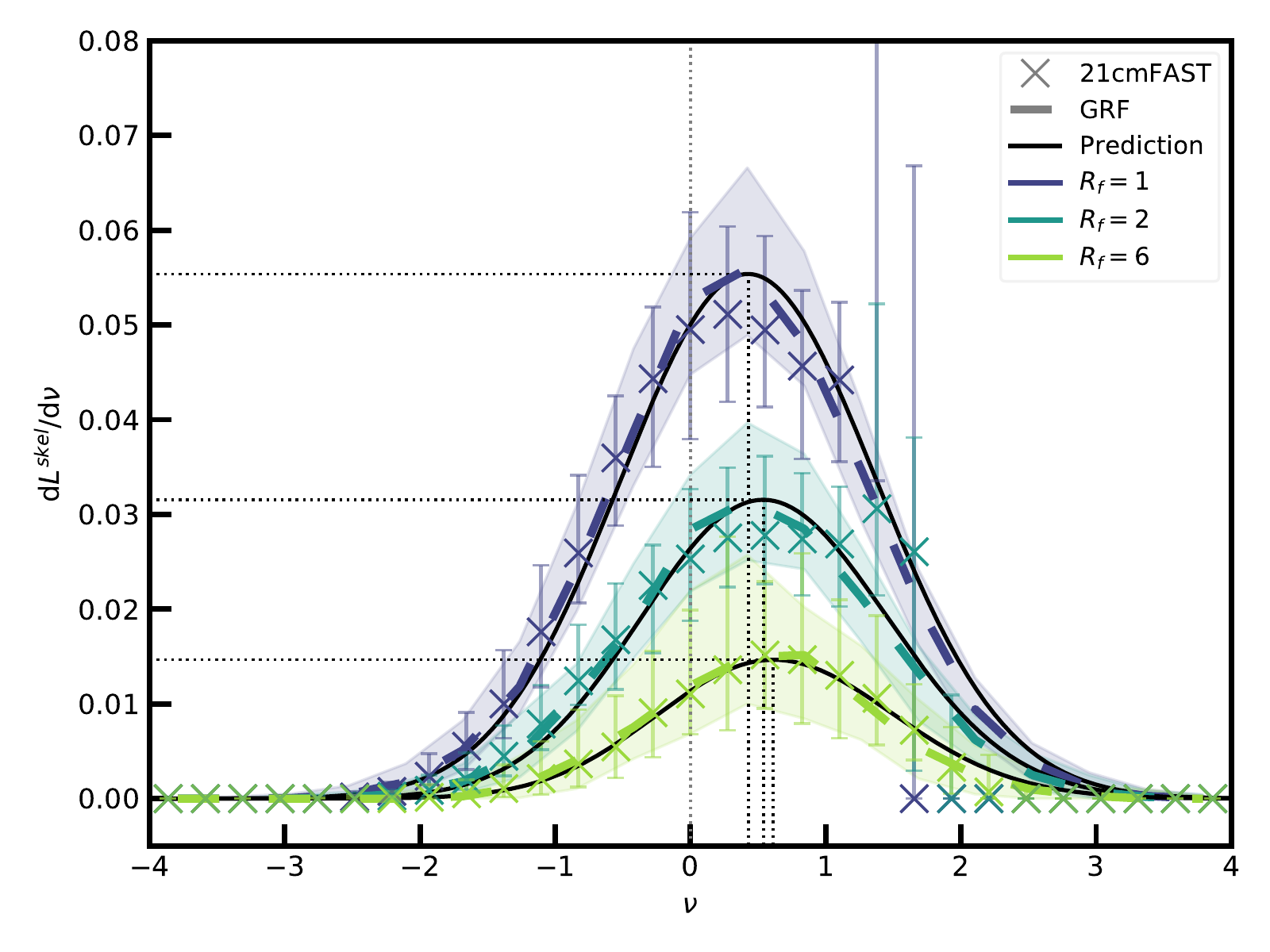}
    \caption{Distribution of the skeleton length of the fields for the different smoothings (in colours). The median of every run is computed for each field. The dashed lines correspond to the GRFs, and the crosses are for the $\cmfast$ reionisation times fields. The black lines are the theoretical predictions. The shaded areas and the error bars represent the dispersion around the median (1\textsuperscript{st} and 99\textsuperscript{th} percentiles) of the GRFs and $t_{\text{reion}}(\Vec r)$ respectively. The black dotted vertical lines represent the average of the predictions. Here, $\nu$ represents the value of the normalised reionisation times.}
    \label{fig:Lskel_21cmFAST_GRFs_treion}
\end{figure}

On Fig. \ref{fig:Map_histgrad_21cmFAST_GRFs_treion}, we show the 2D histograms for $t_{\text{reion}}(\Vec r)$ on the first row and the GRFs on the second row. Each column represents a different gaussian smoothing with $R_f\in\{1,2,6\}$ respectively. Figure \ref{fig:QHII_21cmFAST_GRFs_treion} shows the fraction of ionised volume, and Fig. \ref{fig:Lskel_21cmFAST_GRFs_treion}, the PDF of the skeleton length. Both figures show the GRF predictions in black, the $\cmfast$ reionisation times fields in colored crosses, and the GRFs measurements in colored dashed lines. 
In the $\cmfast$ field, there are more imprints of non-gaussianities: the gradient norms decrease dramatically at higher times (i.e. the front velocities increase dramatically). This was already the case in the $\emma$ simulation, but it is much more marked in the present case. For the two smallest smoothings, it also appears in the $Q_{\text{HII}}$ statistic (see Fig. \ref{fig:QHII_21cmFAST_GRFs_treion}), where the gas seems to be totally reionised earlier than the predictions, as well as in the skeleton length PDFs (see Fig. \ref{fig:Lskel_21cmFAST_GRFs_treion}), where at later times, the distributions depart from the predictions. 
In fact, $\cmfast$ does not explicitly model radiation propagation \citep{Zahn2011}, which means that the front velocities are not limited by the speed of light for example. This is not the case in $\emma$, and can explain the small gradient values at late times.

Nevertheless, even if the $\cmfast$ measurements of Figs. \ref{fig:QHII_21cmFAST_GRFs_treion} and \ref{fig:Lskel_21cmFAST_GRFs_treion} are not superimposed to the predictions, they stay in the measurements error bars. For the two smallest smoothing kernels, the $\cmfast$ skeleton length distributions (Fig. \ref{fig:Lskel_21cmFAST_GRFs_treion}) peak at a lower skeleton length value than the $\emma$ distributions (Fig. \ref{fig:Lskel_EMMA_GRFs_treion}). It means that at a given time, the skeleton lengths are smaller in the $\cmfast$ simulation than in the $\emma$ simulation: there are less percolation places in the $\cmfast$ simulation than in the $\emma$ simulation at a given moment. As already mentioned, a shorter skeleton length implies larger patches for a simple lattice model: it would be consistent with the results obtained in \citet{Thelie2022}, where $\cmfast$ patches were usually found to be larger than the ones found in $\emma$ simulation for similar models.

\section{Conclusion \& Perspectives}
\label{sec:conclusion}

In this work, we extract topological statistics from 2D reionisation times (and redshift in App. \ref{app:zreion}) maps coming from an $\emma$ cosmological simulation and a $\cmfast$ semi-analytical simulation (that have approximately the same reionisation history). reionisation times maps contain a wealth of a spatial and temporal information about the reionisation process. 
The fraction of ionised volume (i.e. the filling factor of the $t_{\text{reion}}(\Vec{r})$ map) contains information on the global timing and evolution of the reionisation process. The PDF of the gradients norm map informs us about the radiation fronts velocity. The average isocontour length of the field allows us to follow the percolation process. The critical points distributions inform us about the timing of appearance of the reionisation seeds. The skeleton lengths tell us about the moment, duration and place of the percolation of ionisation fronts.
We also apply the gaussian random field (GRF) theory \citep{Rice1944,Longuet1957,Doroshkevich1970,BBKS,Gay2012} in the context of the EoR to compare GRF predictions to measurements of these statistics on simulations. 
We generate GRFs from a fitted power spectrum of each simulation to check our simulations measurements. 

We have shown that the topological statistics extracted from the $\emma$ and $\cmfast$ reionisation times maps are rather close to the GRF predictions, and even more when the maps are smoothed on larger areas. It means that $t_{\text{reion}}(\Vec{r})$ can be supposed to be gaussian with a good level of approximation, and that we have therefore developed a simple tool that allows us to quickly generate fields related to reionisation. This result is surprising in a context where many other EoR fields have been shown to be highly non-gaussian, such as the 21 cm or the density fields (see, for example, \citet{Mellema2006,Iliev2006,Majumdar2018,Ross2019}).
Now, the major differences between the $\emma$ cosmological simulation and the $\cmfast$ semi-analytical simulation reionisation times fields seem to be caused by the increase of the fronts velocity at the end of the EoR.

The topological statistics applied to reionisation times field, can therefore be used to characterise the evolution of the EoR. The reasonable agreement between GRFs predictions and models measurements also suggests the possibility of generating histories of reionisation on the sky from the simple knowledge of the power spectrum of reionisation times field. Such generated histories would automatically come with a set of topological statistics (number of reionisation seeds, skeleton length, Minkowski functionals, etc.) fully determined by the power spectrum within the framework of GRF theory.
Besides, we show here that the reionisation evolution can be inferred from the power spectrum parameters (or the spectral parameters $R_0$, $R_*$, and $\gamma$) only, as long as the scales are large enough so that the reionisation times field is close to a GRF. Finally, the topological statistics discussed here depend directly on the power spectrum parameters (amplitudes, slopes, characteristic scales) in the gaussian random field approximation. The physics of the propagation of reionisation, presumably encoded in the power spectrum, can be constrained even in situations where the power spectrum cannot be easily estimated, by fitting e.g. peaks, isocontours or skeleton statistics with their gaussian predictions. As such, they can be used to constrain the power spectrum, even in situations where the reionisation times fields suffer from e.g. noise or poor resolution.

However, our studies show that this similarity with GRFs predictions operates on large scales about 8 cMpc/h, i.e. similar to the SKA resolution at these redshifts. We still have small imprints of non-gaussianities on the smaller scales. Indeed, at the end of the EoR, the radiation fronts propagate faster and faster, due to the remaining neutral voids. It makes the process asymmetric with respect to the mean reionisation time, and it is poorly reconstructed with the symmetric theory that is the GRF theory. As the regions that remain to be ionised get smaller and smaller as the EoR ends, this phenomenon stays at small scales, and this velocity increase gets smoothed out with the largest smoothing.
To take into account these asymmetries, we could add non-gaussian terms in our expressions with the Gram-Charlier expansion \citep{Gay2012,Cadiou2020}. Also, it could be interesting to investigate how reduced speed of light approximations \citep{Deparis2019,Ocvirk2019} influence the statistics presented here and could lead to an even better agreement with GRFs predictions.
In any case, our results are probably resolution-dependent and we could verify it with models that have a better resolution.

Finally, the reionisation times or redshifts fields are not directly observable. In the next decade, we will acquire with the SKA observatory 21 maps of the EoR that will have similar resolution to our simulation smoothed with $R_f=6$. That is why, we are creating a method in a future paper (Hiegel et al. in preparation) to reconstruct 2D reionisation redshifts maps from 2D 21 cm maps (that are taken at a given redshift), and from which we can compute the reionisation times maps.
With these maps, we will be able to infer the topological characteristics of the reionisation process as we did here with simulations.

\section*{Acknowledgements}

We thank Christophe Pichon and Corentin Cadiou for their help and advice on the understanding of the GRF theory.
This work was granted access to the HPC resources of CINES and TGCC under the allocations 2020-A0070411049, 2021- A0090411049, and 2022-A0110411049 “Simulation des signaux et processus de l’aube cosmique et Réionisation de l’Univers” made by GENCI.
This research made use of \textsc{tools21cm}, a community-developed package for the analysis of the 21 cm signals from the EoR and Cosmic Dawn \citep{Giri2020}, \textsc{astropy}, a community-developed core Python package for astronomy \citep{AstropyCollab2018}; \textsc{matplotlib}, a Python library for publication quality graphics \citep{Hunter2007}; \textsc{scipy}, a Pythonbased ecosystem of open-source software for mathematics, science, and engineering \citep{Virtanen2020}; \textsc{numpy} \citep{Harris2020} and \textsc{Ipython} \citep{Perez2007}.

\appendix
\section{Reconstruction of reionisation times from 21 cm}
\label{app:hiegel2022}

\begin{figure*}
   \centering
   \includegraphics[width=1\textwidth,trim={2.75cm 0 2.95cm 0},clip]{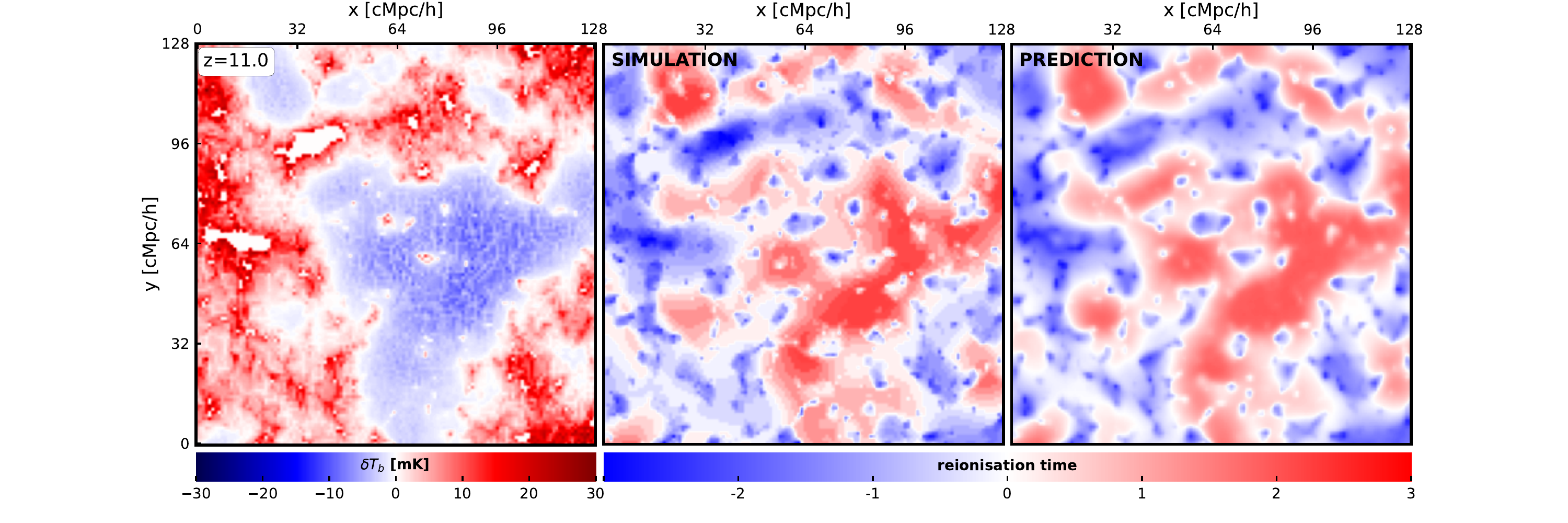}
    \caption{Example of the reconstruction of a 2D reionisation times map from a 2D map of the 21 cm signal taken at a redshift $z=11$. The left and middle panels are the brightness temperature and reionisation times fields generated by a $\cmfast$ simulation. The right panel is the reconstruction of the reionisation times with the neural network that aims at reproducing the true reionisation times map of the middle panel. Both $t_{\text{reion}}(\Vec r)$ maps are dimensionless.}
    \label{fig:ReconstructionsOFTreion}
\end{figure*}

In this paper, we discuss the reionisation times field that informs us about the time at which the gas is reionised at each position. It holds spatial and temporal information about the reionisation process. 
Even though this field is primarily available only via EoR models, we aim at being able to also work on an ‘observed' reionisation times field, from 21 cm maps. 
For example, the SKA will collect the (redshifted) 21 cm signal to produce 2D images on the plane of the sky at many redshifts (or frequencies) along the line of sight. These images will thus contain the differential brightness temperature $\delta T_b$ relative to a background radio temperature, and map the distribution of neutral hydrogen at different redshifts. On the left panel of Fig. \ref{fig:ReconstructionsOFTreion}, we show an example of a $\cmfast$ 21 cm map at a redshift of 11. In Hiegel et al. (in preparation), we aim at reconstructing the reionisation times from this signal and this appendix summarized what has been currently achieved on this objective.

We used a convolutional neural network (CNN) algorithm that can learn and detect complex pattern within images. In particular, we have developed a U-net that takes an image as input and reconstruct another image as an output: in our case the inputs are 2D 21 cm maps at a given redshift (such as the one on the left panel of Fig. \ref{fig:ReconstructionsOFTreion}) and the CNN will learn to construct outputs that will be the closest as possible to the corresponding 2D reionisation times maps (such as the one on the middle panel of Fig. \ref{fig:ReconstructionsOFTreion}). We therefore constructed data sets of fifty $\cmfast$ simulations that have a size of 256 cMpc/h with a resolution of 1 cMpc/h, from which we extract $128\times128$ images. These images are split into a training set on which the CNN will learn to reconstruct $t_{\text{reion}}(\Vec r)$ and a validation set to check its performance. In that paper, we do not smooth the reionisation times maps.

With this U-net, we are able to reconstruct reionisation times maps from observation-like maps with levels of correspondence to the true maps that vary with the observational redshift. For $z\in [8-12]$, 65\% to 96\% of $t_{\text{reion}}(\Vec r)$ signal is well reconstructed. The reconstructed map shown in the right panel of Fig. \ref{fig:ReconstructionsOFTreion} has been obtained with 21 cm maps taken at $z=11$, which is one of the redshift that returns the best results. We can see that in the process, and even for the best reconstructions, the small scales are smoothed out of the predicted $t_{\text{reion}}(\Vec r)$ maps compared to the true ones, which is to be improved in future works. In Hiegel et al. (in preparation), we quantify the performance of the CNN via many diagnostics and we show for example that we can extract an ionisation history that is consistent with the ones from the $\cmfast$ simulations. It means that we can extract information about the evolution of the reionisation process and its topology only from a 21 cm map obtained at a single redshift.

\section{Calculation of the moment of a field and its derivatives from a given power spectrum}
\label{app:std} 

In this work, we use the spectral moments $\sigma_i$ (for $i\in \mathbb{N}$) of the field of interest, in order to normalise our fields, or because they appear in spectral parameters (defined in Eq. \ref{eq:spectralParameters}). These moments are only defined by the power spectrum of the field. The zeroth order moment of a field is simply the standard deviation of the field, the first order moment is the standard deviation of the first derivative of the field, and so on, as written in Eq. \ref{eq:firstSigma}. 

The moments can be expressed as follows \citep{BBKS,Pogosyan2009,Gay2011}:
\begin{equation}
    \sigma_i^2 = \frac{ 2\pi^{\frac{d}{2}} }{ \Gamma\left(\frac{d}{2}\right) } \int_0^\infty k^{2i} \mathcal{P}_k k^{d-1} dk,
\end{equation}
where $i\in \mathbb{N}$ corresponds to the number of derivation of the field, and $d$ is the dimension of the field (in our case, $d=2$). In our case, we are interested in the power spectra of the reionisation field, which are defined in Eq. \ref{eq:Pk}, and which have two slopes in logarithmic scales (in his thesis, \citet{Gay2011} do the calculation for a power spectrum with one slope in logarithmic scales). To do so, we need the gamma functions, defined below:
\begin{equation}
\gamma(a,x) = \int_0^x t^{a-1} e^{-t} dt \quad \text{and} \quad
\Gamma(a,x) = \int_x^\infty t^{a-1} e^{-t} dt.
\label{eq:gammaFunctions}
\end{equation}
The integral within the moments can then be separated into two integral where the cut is at the threshold $k_{\text{thresh}}$  separating the two parts of the power spectrum:
\begin{equation}
    \begin{split}
        \sigma_i^2 & = \frac{ 2\pi^{\frac{d}{2}} }{ \Gamma\left(\frac{d}{2}\right) } \Biggl[ \int_0^{k_{\text{thresh}}} k^{2i} \mathcal{P}_k k^{d-1} dk + \int_{k_{\text{thresh}}}^\infty k^{2i} \mathcal{P}_k k^{d-1} dk \Biggr] \\
        & \begin{split}
            = \frac{ A_2 2\pi^{\frac{d}{2}} }{ \Gamma\left(\frac{d}{2}\right) } & \Biggl[ \frac{A_1}{A_2} \int_0^{k_{\text{thresh}}} k^{d-1+2i+n_1} e^{-2R_f^2k^2} dk \\
            & + \int_{k_{\text{thresh}}}^\infty k^{d-1+2i+n_1} e^{-2R_f^2k^2} dk \Biggr] \\
        \end{split} \\
        & \begin{split}
            = \frac{ A_2 2\pi^{\frac{d}{2}} }{ \Gamma\left(\frac{d}{2}\right) } & \Biggl[ \frac{A_1}{A_2} \left(\frac{1}{\sqrt{2}R_f}\right)^{d+2i+n1} \int_0^{2R_f^2k_{\text{thresh}}^2} k^{\frac{d+n_1}{2}-1} e^k dk \\
            & + \left(\frac{1}{\sqrt{2}R_f}\right)^{d+2i+n2} \int_{2R_f^2k_{\text{thresh}}^2}^\infty k^{\frac{d+n_2}{2}-1} e^k dk \Biggr]. \\
        \end{split} \\
    \end{split}
\end{equation}
Now using the gamma functions defined above, we have the following expression of the spectral moments:
\begin{equation}
\begin{split}
    \sigma_i^2 = \frac{ A_2 2\pi^{\frac{d}{2}} }{ \Gamma\left(\frac{d}{2}\right) } & \Biggl[ \frac{A_1}{A_2} \left(\frac{1}{\sqrt{2}R_f}\right)^{d+2i+n1} \gamma\Biggl(\frac{d+n1}{2}+i, 2R_f^2k_{\text{thresh}}^2\Biggr)  \\
    & + \left(\frac{1}{\sqrt{2}R_f}\right)^{d+2i+n2} \Gamma\Biggl(\frac{d+n2}{2}+i, 2R_f^2k_{\text{thresh}}^2\Biggr) \Biggr].
\end{split}
\end{equation}
The moments $\sigma_i^2$ have the same units as the power spectrum, but it is worth to mention that the spectral parameters $R_0$, $R_*$ and $\gamma$ remains dimensionless. Note also that they are only dependent on the dimension and the power spectrum parameters: $\sigma_i^2 = \sigma_i^2(d,A_1,n_1,A_2,n_2,R_f)$.

\section{Calculation of the PDF of the gradient norm of a field}
\label{app:PDFgradients}

Our fields of interest being gaussian, we remind that their PDF can be written as follows:
\begin{equation}
    P(\vec x) d^n\vec x = \frac{1}{(2\pi)^{n/2} \cdot \det(C)^{\frac{1}{2}}} \exp\left(-\frac{1}{2} \vec x \cdot C^{-1} \cdot \vec x\right) d^n\vec x,
\end{equation}
where $\vec x$ is a n-D vector function of the position and $C = \left<\vec x \otimes \vec x\right>$ is the covariance matrix. To compute a PDF depending on the field and its first derivative, we use $\vec x = \textvec{x\\x_1\\x_2}$. 

The covariance matrix of a 2D field $F$ is the following:
\begin{equation}
\begin{split}
    C & =
    \begin{pmatrix}
        \left< F^2 \right> & \left< F\ \nabla_1 F \right> & \left< F\ \nabla_2\ F \right> \\
        \left< \nabla_1 F\ F \right> & \left< \left(\nabla_1 F\right)^2 \right> & \left< \nabla_1 F\ \nabla_2 F \right> \\
        \left< \nabla_2 F\ F \right> & \left< \nabla_1 F\ \nabla_2 F \right> & \left< \left(\nabla_2 F\right)^2 \right>  
    \end{pmatrix} \\
    & =
    \begin{pmatrix}
        \sigma_0^2 & 0 & 0 \\
        0 & \frac{1}{2}\sigma_1^2 & 0 \\
        0 & 0 & \frac{1}{2}\sigma_1^2 
    \end{pmatrix}. \\
\end{split}
\end{equation}
Using the normalised variables $x$, $x_1$, and $x_2$, the covariance matrix becomes:
\begin{equation}
C = 
\begin{pmatrix}
        1 & 0 & 0 \\
        0 & \frac{1}{2} & 0 \\
        0 & 0 & \frac{1}{2}
\end{pmatrix}. 
\end{equation}

All the components of the PDF are now known, and after calculations, it is expressed as follows (and Eq. \ref{eq:PDFgradnorm} is retrieved):
\begin{equation}
    P(x,x_1,x_2)dxdx_1dx_2 = \frac{2}{(2\pi)^{3/2}} e^{-\left(\frac{1}{2} x^2 + x_1^2 + x_2^2\right)} dxdx_1dx_2.
\end{equation}

As we are interested in PDFs only depending the first derivative of $F$, an integration on the field values $x$ is done (thanks to the Gaussian integral $\int_{-\infty}^{\infty} e^{-\alpha y^2}dy = \sqrt{\frac{\pi}{\alpha}}$):
\begin{equation}
    P(x_1,x_2)dx_1dx_2 = \frac{1}{\pi} e^{-\left(x_1^2+x_2^2\right)}dx_1dx_2.
\end{equation}

Moreover, we are interested in the norm of the gradient of the field, that is why we make a change of variable and introduce $w^2 = x_1^2+x_2^2$. It is, in fact, a change of variables in 2D polar coordinates:
\begin{equation}
    \forall \theta\in[0,2\pi], \quad
    \begin{cases}
        x_1 = w\cos(\theta), \\
        x_2 = w\sin(\theta). \\
    \end{cases}
\end{equation}
With this change of variable, and as the PDF is independent on the introduced angle $\theta$, we can write:
\begin{equation}
    P(x_1,x_2)dx_1dx_2 = P(w,\theta)wdwd\theta = 2\pi P(w)wdw = 2w e^{-w^2} dw,
\end{equation}
with a rewritten PDF depending only on the norm of the gradient of the field:
\begin{equation}
    P(w) =  \frac{1}{\pi} e^{-w^2}.
\end{equation}

\section{reionisation redshifts field analyses}
\label{app:zreion}

The reionisation times and redshifts fields are related with the same expression linking time and redshift:
\begin{equation}
    z(t) = \frac{1}{a(t)}-1,
\end{equation}
with $a$ the scale factor. With this definition, they have opposite monotonies, which has a consequence on a topological study. Indeed, the two fields have small differences: $t_{\text{reion}}(\Vec r)$ increases more rapidly as $z_{\text{reion}}(\Vec r)$ decreases, and it causes some distinctions in gaussianity analyses.

The aim of this appendix is only to present a few results on the reionisation redshifts field ($z_{\text{reion}}(\Vec r)$) that presents some differences to the reionisation times field. To remain brief, we focus on the $\emma$ $z_{\text{reion}}(\Vec r)$. From the same simulation described in Sect. \ref{sec:EMMAsimu}, we can extract a hundred slices of the reionisation redshifts field, that are also smoothed with a gaussian kernel of standard deviation $R_f\in\{1,2,6\}$, and normalised as below:
\begin{equation}
    z_{\text{reion}} = \frac{z_{\text{reion}}^*-\overline{z_{\text{reion}}^*}}{\sigma_0^{z_{\text{reion}}^*}}, 
\end{equation}
with $\overline{z_{\text{reion}}^*}$ the mean of each field. $\sigma_0^{z_{\text{reion}}^*}$ is the expected standard deviation of the field. The average and standard deviation of the reionisation redshifts fields are given in Tab. \ref{table:Simu_avg_and_std}. The power spectrum of the reionisation redshifts field is also fitted as described in Sect. \ref{sec:EMMAsimu}, and the resulting parameters are shown in Tab. \ref{table:powerSpectraParams}. 

We generate again a hundred of GRFs with the proper power spectrum to compare $z_{\text{reion}}(\Vec r)$ to them. They are also smoothed and normalised as described in Sect. \ref{sec:GRFsimu}.

The resulting predictions for the reionisation times field are also calculated for the reionisation redshifts field and shown below. The expressions are slightly modified for $z_{\text{reion}}(\Vec r)$, changing some integrals limits or signs.

\vspace{11pt}
\noindent \textbf{\large{Filling factor}}

From Eq. \ref{eq:PDFFieldValues}, we can calculate the fraction of ionised volume of the reionisation redshifts the same way as the one of the reionisation times. Here, the number of values that has a redshift higher than the threshold is equivalent to the number of cells that has already reionised. It corresponds then directly to the fraction of ionised volume $Q_{\text{HII}}$, as written below: 
\begin{equation}
    Q_{\text{HII}}(\nu) = \int_\nu^\infty P(x)dx = \frac{1}{2} \erfc\left(\frac{\nu}{\sqrt{2}}\right).
\end{equation}

\vspace{11pt}
\noindent \textbf{\large{Filling factor of the gradient norm}}

As in the previous section, the filling factor of the gradient norm of redshift fields can be obtained from the joint PDF of Eq. \ref{eq:PDFgradnorm}, and is defined as follows:
\begin{equation}
    f(\nu) = \int_{\nu}^\infty 2\pi P(w)wdw = e^{-\nu^2}.
\label{eq:ff_grad_z}
\end{equation}

\vspace{11pt}
\noindent \textbf{\large{Isocontours length}}

As it is symmetric, the isocontours length is the same one for the reionisation times and redshift fields, for which the expression is reminded below:
\begin{equation}
    \mathcal{L}(\nu) = \frac{1}{2\sqrt{2}R_0} e^{-\frac{1}{2}\nu^2}.
\end{equation}

\vspace{11pt}
\noindent \textbf{\large{Distribution of the maxima}}

The maxima of the reionisation redshifts slices and the GRFs can be extracted with $\disperse$, as explained is Sect. \ref{sec:ReionisationSeedCount}. Their distribution can be theoretically calculated the same way as for the minima (as described in Sect. \ref{sec:distrib_cp}). We obtain the following distribution, where only some signs have changed:
\begin{equation}
\footnotesize
\begin{split}
    \frac{\partial n_{\text{max}}}{\partial \nu}   & = \frac{ \exp\left({\frac{-\nu^2}{2}}\right) }{ \sqrt{2\pi}R_*^2 } \left[1 + \erf{ \left(\frac{\gamma \nu}{\sqrt{2(1-\gamma^2}} \right) }\right] K_1(\nu,\gamma) \\
    & + \frac{ \exp\left(\frac{-3\nu^2}{6-4\gamma^2}\right) }{ \sqrt{2\pi(1-\frac{2}{3}\gamma^2)}R_*^2 } \left[1 + \erf{ \left(\frac{\gamma \nu}{\sqrt{2(1-\gamma^2)(3-2\gamma^2)}} \right)} \right] K_2 \\
    & + \frac{ \exp\left(\frac{-\nu^2}{2(1-\gamma^2)}\right) }{ \sqrt{2\pi(1-\gamma^2)}R_*^2 } \left[1 + \exp\left( \frac{\nu^2}{2(1-\gamma^2)} \right) \right] K_3(\nu,\gamma). \\
\end{split}
\label{eq:dn_dnu_z}
\end{equation}

\vspace{11pt}
\noindent \textbf{\large{Distribution of the antiskeleton length}}

As the reionisation redshifts field has an opposite monotony compared to the reionisation times field, the skeleton of $t_{\text{reion}}(\Vec r)$ is equivalent to the antiskeleton of $z_{\text{reion}}(\Vec r)$. The antiskeleton joins minima together passing through saddle points, and it can also be extracted from the fields thanks to $\disperse$. The distribution of the antiskeleton lengths are calculated the same way as the skeleton length distribution. \citet{Gay2012} informs us that it results in the same expression as the one of the skeleton length (see Eq. \ref{eq:Lskel}) but with $\nu$ that becomes $-\nu$, as follows:

\begin{equation}
\footnotesize
\begin{split}
    \frac{\partial L^\text{skel}}{\partial \nu} & = \frac{1}{\sqrt{2\pi} R_*} \exp^{-\frac{\nu^2}{2}} \Biggr[ \frac{1}{8\sqrt{\pi}} (\sqrt{\pi} - 2 \gamma \nu) \left( 1 + \erf\left( \frac{- \gamma \nu}{\sqrt{2(1-\gamma^2)}} \right) \right) \\
    & +  \frac{\sqrt{1-\gamma^2}}{2\sqrt{2}\pi} \exp\left( - \frac{\gamma^2\nu^2}{2(1-\gamma^2} \right) \Biggr]. \\
\end{split}
\label{eq:Lskel_z}
\end{equation}

\vspace{11pt}
\noindent \textbf{\large{Results}}

\begin{figure*}
   \centering
   \includegraphics[width=1\textwidth]{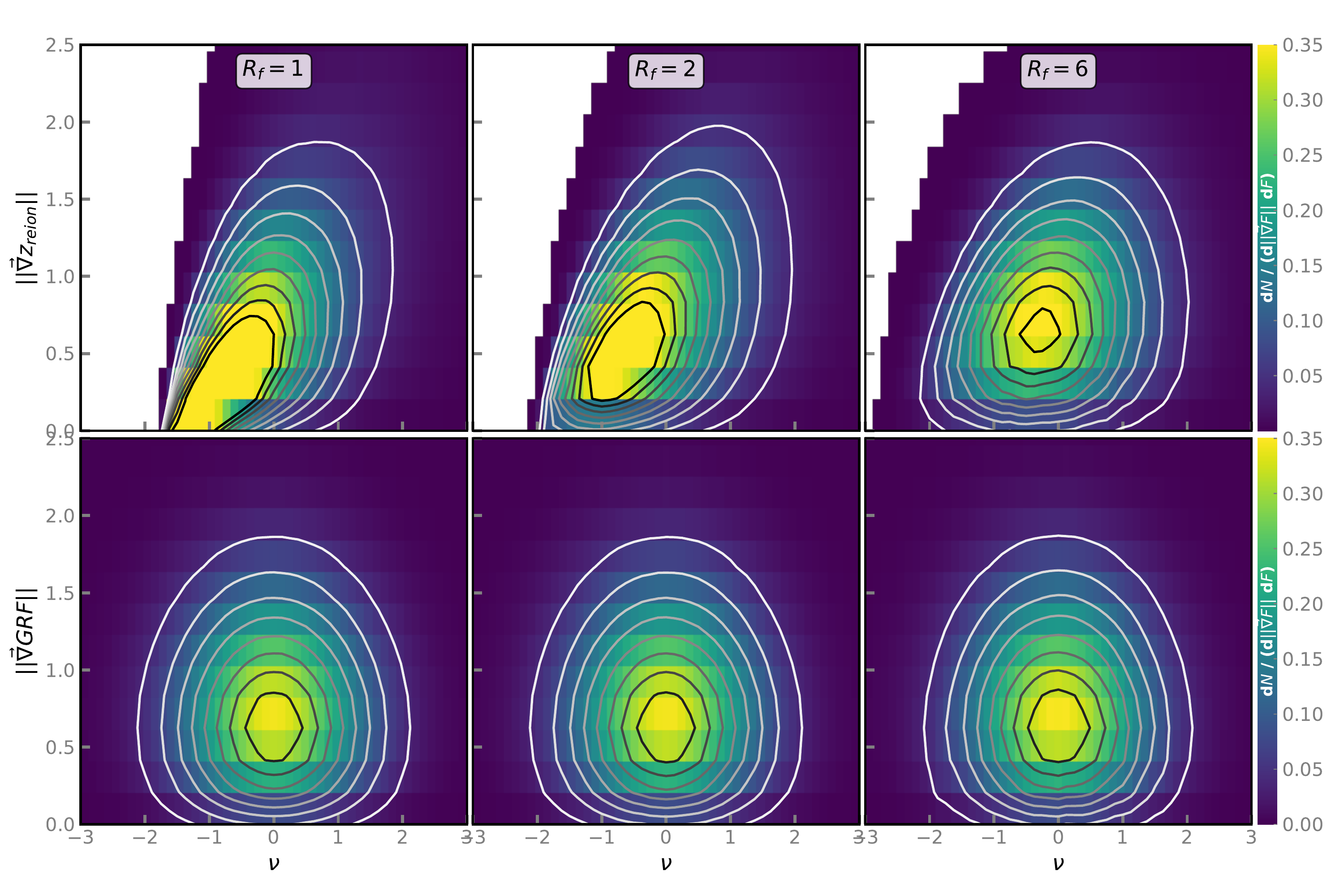}
    \caption{2D PDFs of the gradient norms with respect to the values of the fields of every run for each field and different smoothings ($R_f\in\{1,2,6\}$, see each column). The first row corresponds to the $\emma$ reionisation redshifts, and the second row to their corresponding GRFs. The gray-scale lines are the isocontours of the histograms. Here, $\nu$ represents the value of the normalised reionisation redshifts.}
    \label{fig:Map_histgrad_EMMA_GRFs_zreion}
\end{figure*}

\begin{figure}
   \centering
   \includegraphics[width=0.5\textwidth]{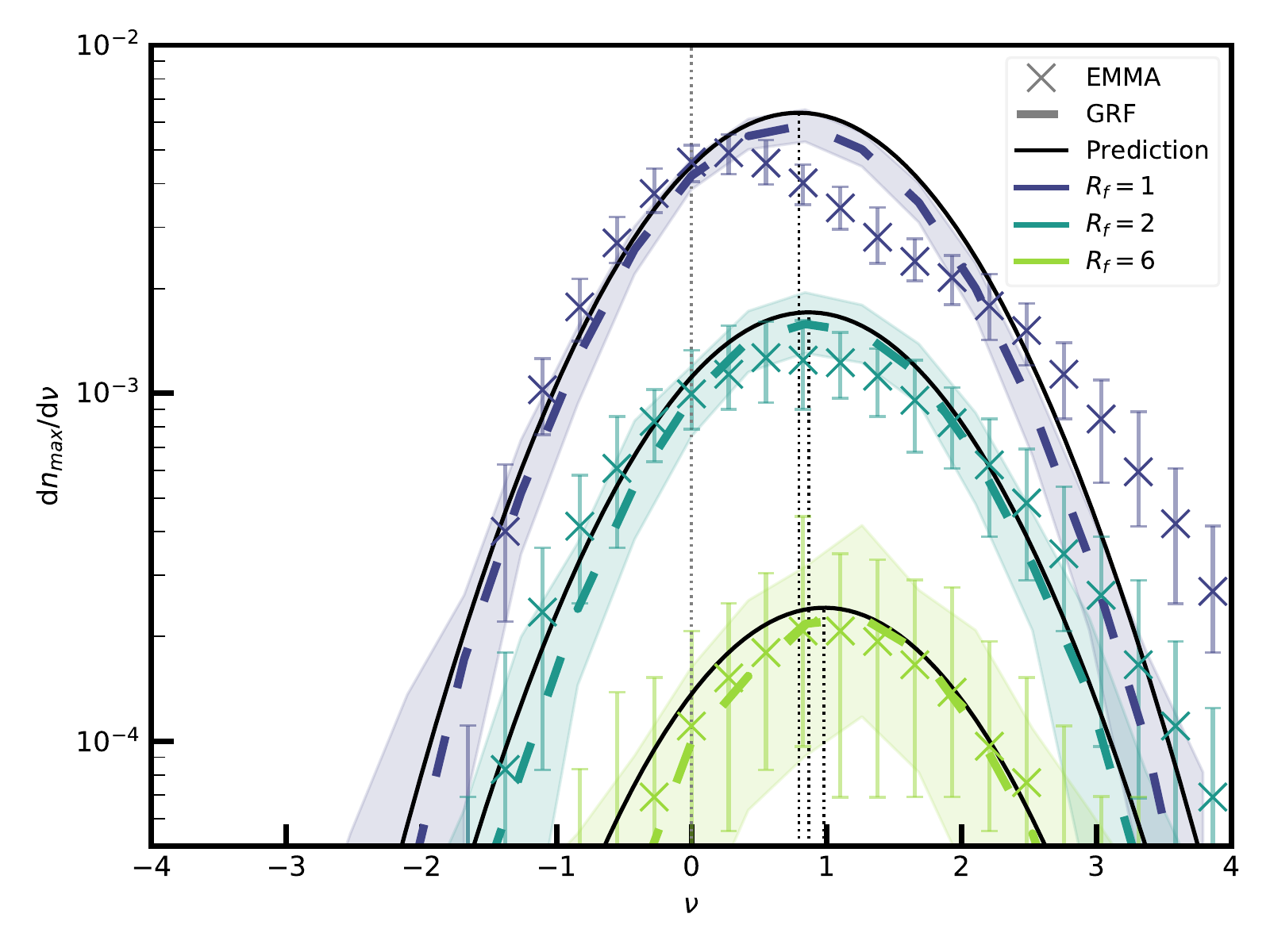}
    \caption{Distribution of the critical points of the fields for the different smoothings (in colours). The median of every run is computed for each field. The dashed lines correspond to the GRFs, and the crosses are for the $\emma$ reionisation redshifts fields. The black lines are the theoretical predictions. The shaded areas and the error bars represent the dispersion around the median (1\textsuperscript{st} and 99\textsuperscript{th} percentiles) of the GRFs and $t_{\text{reion}}(\Vec r)$ respectively. The black dotted vertical lines represent the average of the predictions. Here, $\nu$ represents the value of the normalised reionisation redshifts.}
    \label{fig:distrib_cp_EMMA_GRFs_zreion}
\end{figure}

In this short result section, we only show figures that highlight the discrepancies between the reionisation times and redshifts fields. We begin with the 2D histograms of the gradient norm of $z_{\text{reion}}(\Vec r)$ versus $z_{\text{reion}}(\Vec r)$ for the $\emma$ simulation in Fig. \ref{fig:Map_histgrad_EMMA_GRFs_zreion}. On the first row, there are the cosmological fields, and on the second row, there are the GRFs. There is again a ‘symmetrisation' when the size of the gaussian kernel increases (i.e. $R_f$ increases), although it is less convincing than the reionisation times field (see Fig. \ref{fig:Map_histgrad_EMMA_GRFs_treion}).
On Fig. \ref{fig:distrib_cp_EMMA_GRFs_zreion}, we show the maxima PDF of the $\emma$ reionisation redshifts fields (with the crosses), of the corresponding GRFs (with the dashed lines), and the prediction (in black). Comparing them to the minima PDF of the $\emma$ reionisation times field (see Fig. \ref{fig:distrib_cp_EMMA_GRFs_treion}), we can see that the $t_{\text{reion}}(\Vec r)$ minima PDFs are closer to the GRF ones than the $z_{\text{reion}}(\Vec r)$ maxima PDFs, which again shows that $t_{\text{reion}}(\Vec r)$ is more gaussian than $z_{\text{reion}}(\Vec r)$. 
With these PDFs, we can again see the effect of the non-linear relation between $z_{\text{reion}}(\Vec r)$ and $t_{\text{reion}}(\Vec r)$, which affects the x-axis. Indeed, for the smallest smoothing, the difference between the time and redshift of reionisation is evident: $z_{\text{reion}}(\Vec r)$ underestimates the number of critical points (that are reionisation seeds) with respect to $t_{\text{reion}}(\Vec r)$. 
These figures globally show that the reionisation times field is more gaussian than the reionisation redshifts field, which is due to the non-linear relation between time and redshift, and which affects all the statistics studied in this paper.

%
\bibliographystyle{aa} 
\bibliography{biblio} 
%

\end{document}